\documentclass{aa}  
\usepackage{natbib}
\bibpunct{(}{)}{;}{a}{}{,} 
\usepackage{graphicx}
\usepackage{float}
\usepackage{xcolor}
\usepackage{graphicx}
\usepackage{dcolumn}
\usepackage{bm}
\usepackage{romannum}
\usepackage{amsmath}
\usepackage{graphicx}
\usepackage{xcolor}
\usepackage[caption=false]{subfig}
\usepackage{booktabs}   
\usepackage{relsize}

\usepackage{txfonts}

\begin{document}

\title{Galaxy formation with wave/fuzzy dark matter: The core-halo structure and the solitonic imprint}

   \author{A. Pozo
          \inst{1,2}
          \and
          R. Emami\inst{3}
          \and
          P. Mocz\inst{4,5}
          \and
          T. Broadhurst\inst{1,2,6}
          \and
          L. Hernquist\inst{3}
          \and
          M. Vogelsberger\inst{11}
          \and
          R. Smith\inst{3}
          \and
          G. Tremblay\inst{3}
          \and
          R. Narayan\inst{3,12}
          \and
          J. Steiner\inst{3}
          \and
          J. Grindlay\inst{3}
          \and
          G. Smoot\inst{2,7,8,9,10}
          }

  \institute{University of the Basque Country UPV/EHU,Department of Theoretical Physics, Bilbao, E-48080, Spain\\
              \email{alvaro.pozolarrocha@bizkaia.eu; tom.j.broadhurst@gmail.com;}
         \and
             DIPC, Basque Country UPV/EHU, E-48080 San Sebastian, Spain
         \and
             Center for Astrophysics $\vert$ Harvard $\&$ Smithsonian, 60 Garden Street, Cambridge, MA 02138, USA
          \and
             Department of Astrophysical Sciences, Princeton University, 4 Ivy Lane, Princeton, NJ, 08544, USA
          \and
             Lawrence Livermore National Laboratory, 7000 East Ave, Livermore, CA 94550, USA
         \and
             Ikerbasque, Basque Foundation for Science, Bilbao, E-48011, Spain
         \and
             Hong Kong University of Science and Technology, Institute for Advanced Study and Department of Physics, IAS TT $\&$ WF Chao Foundation Professor,  Hong Kong
         \and
             Energetic Cosmos Laboratory, Nazarbayev University, Nursultan, Kazakhstan
         \and
             Paris Centre for Cosmological Physics, APC, AstroParticule et Cosmologie, Universit\'{e} de Paris, CNRS/IN2P3, CEA/lrfu,Universit\'{e} Sorbonne Paris Cit\'{e}, 10, rue Alice Domon et Leonie Duquet,	75205 Paris CEDEX 13, France  Emeritus
        \and
             Physics Department $\&$ LBNL, University of California at  Berkeley CA 94720 {\it Emeritus}
        \and
             Department of Physics, Kavli Institute for Astrophysics and Space Research, Massachusetts Institute of Technology, Cambridge, MA 02139, USA
        \and
             Black Hole Initiative at Harvard University, 20 Garden Street, Cambridge, MA 02138, USA
             }

  \abstract{
 Dark matter-dominated cores have long been claimed for the well-studied local group dwarf galaxies. More recently, extended stellar halos have been uncovered around several of these dwarfs through deeper imaging and spectroscopy. Such core-halo structures (inner flat core and a characteristic $r^{-3}$ asymptotic outer halo profile) are not a feature of conventional cold dark matter (CDM). In contrast, smooth and prominent dark matter cores are predicted for wave/fuzzy dark matter ($\psi$DM). The question arises as to what extent the visible stellar profiles should reflect this dark matter core structure. Here we compare cosmological hydrodynamical simulations of CDM, “WDM” (model used as a proxy for $\psi$DM) $\&$ $\psi$DM, aiming to predict the stellar profiles for these three DM scenarios. We show that cores surrounded by extended halos are distinguishable for $\psi$DM, where the stellar density is enhanced in the core due to the presence of the relatively dense soliton. Our analysis demonstrates that, in our simulations, a distinctive core-halo structure does not appear in the case of CDM in  the DM, gas, or stars. Whereas we do find a core-halo transition for DM, gas, and stars for $\psi$DM, and the scale of this transition is in line with the predicted core radius set by the soliton scale anticipated for the adopted boson mass  of 2.5$\times10^{-22}$eV. The presence of a core-halo structure in the stellar profile for Galaxy 1 for $\psi$DM is visible for the most massive and the first galaxy to form in the simulation. Clearly, further simulations are needed to establish how strict  this possible relationship is between the DM and stellar core-halo profile as a potential observational discriminator. Furthermore, we observe the anticipated asymmetry for $\psi$DM due to the
soliton’s motion (jumping and random walk), a distinctive characteristic not found in the symmetric distributions of stars in the warm and CDM models.}


   \maketitle

\section{Introduction}

Dark matter is understood to be nonrelativistic, i.e., cold,  as required to form galaxies gravitationally and to explain the spectrum of CMB fluctuations in detail \citep{Planck:2018}. However, the standard heavy particle interpretation for cold dark matter (CDM) faces a stringent laboratory absence of any new particle signature \citep{Aprile:2018,CMS:2020} and several inconsistencies have emerged between CDM predictions and the properties of dwarf galaxies \citep{Moore:1994,deBlok:2010,Marsh:2015,Klypin:1999, Safarzadeh:2021}. Warm dark matter (“WDM”) is defined to be relativistic initially with a kiloelectron volt-scale mass, the 1.4keV mass scale used here for the “WDM” simulation induces the power spectrum suppression, due to early relativistic free streaming, but becomes nonrelativistic later via the cosmological expansion, so that the “WDM” simulation is closely approximated by  N-body simulations. The “WDM” simulations analyzed here were performed using a $\psi$DM initial power spectrum, without including the dynamical quantum potential. This approximation reproduces the small-scale cutoff expected for a thermal relic “WDM” particle of mass 1.4 keV, corresponding to the boson mass of $2.5\times10^{-22}$ eV used in the $\psi$DM simulations of \citet{Mocz:2020}. Alternatively, for very light bosons of $\simeq 10^{-22}eV$, an inherently nonrelativistic possibility of dark matter as a Bose-Einstein condensation \citep{Widrow:1993,Hu:2000}, originally termed fuzzy dark matter (FDM),  emerged from the first simulations in this context, which predict prominent solitonic standing wave cores surrounded by pervasive interference on the de Broglie wavelength scale within all galaxies and filaments \citep{Schive:2014,Mocz:2017,Veltmaat:2018,Hui:2017}. This was then called wave/fuzzy dark matter ($\psi$DM) \citep{Schive:2014,Hui:2021}.
While $\psi$DM predicts  large-scale structure that are almost identical to CDM in terms of the network of filaments and clusters \citep{Schive:2014,Mocz:2017}, the small-scale structure of DM halos is very different in the $\psi$DM paradigm, as DM cannot be confined to less than the de Broglie scale. Consequently, in $\psi$DM, with a light boson mass of $\simeq 10^{-22}eV$, the dwarf galaxy formation is suppressed and there is also a prominent solitonic core \citep{Schive:2014,Schive:20142,Mocz:2017,Veltmaat:2018,Niemeyer:2020} in the galaxy.  

On these scales, self-gravity balances the effective pressure from the uncertainty principle in the ground state at the de Broglie wavelength. Crucially, smaller galaxies are predicted to have wider cores of lower density because the soliton is larger at lower momentum, which we test here.

Many independent claims of dark matter and stellar cores in deep images and spectroscopy are now available for the well-studied Local Group dwarf spheroidal galaxies, dSph, providing one of the main sources of tension for CDM for which the lack of any predicted cores reflects the scale-free nature of the simplest collisionless DM scenario. More recently the presence of extended stellar halos has been uncovered in several local dwarfs, independently, via deep imaging and spectroscopy \citep{Chiti:2021, Collins:2021}. A joint analysis of the available data now suggests that such a core-halo structure with a sharp density transition between the core and the halo is a common feature of  dSph galaxies and also of the smaller new class of ultra-faint galaxies (UFGs) \citep{Pozo:2020,Pozo:2023}. This core-halo behavior \citep{Chan:2020b} motivates the examination of hydrodynamical simulations with a comparison of the differing DM contenders as a possible distinguishing feature.

For cold dark matter (CDM), the scale-free formation of DM structures is predicted to extend to arbitrarily small scales, reaching lower halo masses with higher DM concentrations. In contrast, small-scale suppression of the power spectrum is inherent to both $\psi$DM and “WDM”. In the case of $\psi$DM, this suppression limits the minimum scale of structure to above the de Broglie length, which is determined by the boson mass. Consequently, DM structures below $\simeq 10^{9} M_{\odot}$ are suppressed for a choice of boson masses of $10^{-22}$ eV, motivated by the observed $\simeq 0.3$ kpc scale of dwarf galaxy cores. This limitation has a direct consequence for the formation timescales between CDM and $\psi$DM, and results in a delay in galaxy formation for $\psi$DM compared to CDM. In both cases, we observe that the formation of the first  filamentary structures occur earlier in CDM, where they are comprised of low-mass  subhalos. In contrast, in $\psi$DM/“WDM” there is no fragmentation along the filaments, due to the small-scale cutoff of the power spectrum. Instead, dark matter is distributed more continuously along these filaments {\citep{May:2023,Dome:2023}.

The main difference between  “WDM” and $\psi$DM is that “WDM” exhibits a clear caustic structure in the distribution of dark matter, while $\psi$DM is characterized by wave interference patterns that manifest as fringes in simulations. These fringes are now understood to generate turbulent density structures within halos, resulting from the superposition of numerous plane waves encoding the velocity dispersion in the halo \citep{Mocz:2020}. In $\psi$DM these filaments persist for longer periods as the gravitational attraction from the early formation of the first halos in cold dark matter (CDM) tends to disrupt them earlier. Moreover, the prolonged lifetimes of these filaments in wave/fuzzy dark matter favor the formation of a greater number of stars within them, leading to significant differences in the location and extent of stellar profiles in CDM and $\psi$DM galaxies. This also causes baryonic objects to appear more diffuse or smoothed compared to CDM \citep{Mocz:2019, Mocz:2020}, indicating that they could serve as excellent tracers (as visible objects that could be used to observe the underlying DM structures) of dark matter in $\psi$DM. On the scale of solitons, the quantum pressure in $\psi$DM can become sufficiently strong to counteract the self-gravity of the dark matter. This results in the formation of prominent solitonic standing waves at the base of each virialized potential, characterized by a spherical soliton core of a few kiloparsecs at the center of every dark matter halo \citep{Schive:2014}, in contrast to the much denser cusps observed in CDM \citep{Navarro:1996}.

It is remarkable how well the soliton profile of $\psi$DM aligns with the empirically well-established Burkert profile, as demonstrated in the extensively studied Fornax dwarf spheroidal galaxy \cite{Schive:2014}. However, the scaling of the core radius with galaxy mass reported by \cite{Burkert:2020} from profile fitting to low- and intermediate-mass galaxies exhibits a slope that is opposite in sign to that predicted by $\psi$DM, as highlighted by \cite{Burkert:2020}. Moreover, the relatively large cores inferred for  intermediate-mass galaxies are based on HI gas dynamics. Noncircular gas motions have been recognized as a significant contribution to underestimated rotation curves \citep{Oman:2019,Downing:2023}, together with resolution limitation for HI \citep{Swaters:2000} and the finite thickness of HI disks that all act to bias downward HI-based rotation speed measurements, as  demonstrated using CDM-based hydro simulations with spurious large cores \citep{Roper:2023}. It is important to examine the claimed scaling of galaxy mass and core radius with independent stellar-based dynamics, though this remains challenging due to the typically low stellar surface brightnesses.

In dSph galaxies it is known that stellar motions are dominated by the dark matter gravitational potential, and so we expect some visible differences for stars in these galaxies depending on the nature of dark matter. In addition, the formation and early evolution of galaxies and filaments is also  sensitive to the nature of DM, arising in particular from the suppressed form of the high-k power spectrum for the  “WDM”--$\psi$DM models. The wave behavior and the quantum pressure of $\psi$DM provide different virialized DM halo structures and prominent solitonic cores, unlike the CDM--WDM cuspy profiles (without inner cores) expectations\citep{Navarro:1996} (even smooth cores of  “
WDM are claimed  \citep{Lovell:2014, Maccio:2012,Bode:2001}), $\psi$DM simulations of merging DM halos clearly show the formation of soliton structures at the de Broglie wavelength scale \citep{Schive:2014, Mocz:2017, Schwabe:2016}, and are prominent in terms of the core density, lying above the surrounding DM halo. This is in contrast to the early fragmentation of filaments and cuspy halos in CDM \citep{Mocz:2017,Mocz:2019,Mocz:2020}. \cite{Mocz:2019, Mocz:2020} postulated that baryonic feedback may  not have a significant effect in halos of $10^{9} M_{\odot}$ to $10^{10} M_{\odot}$ for redshifts $z>6$, in the sense of being unable to soften the cuspy profiles of CDM to produce cores. Consequently,  CDM should not  be able to explain the origin of the claimed cores from the dynamical studies of dSph galaxies \citep{Read:2005,Amorisco:2013}.

In this paper we compare the galaxy formation process between CDM, “WDM”, and $\psi$DM, in the high-resolution cosmological simulations by \cite{Mocz:2019,Mocz:2020,Mocz:2017}. Here we consider the distribution of stars modeled in these simulations, where in the “WDM” and $\psi$DM contexts these simulations have already reported stars to be born along dense dark matter filaments, tracing dark matter, and highlighted as ``a smoking gun'' signature of $\psi$DM/“WDM” \citep{Mocz:2019}. Here we  further examine these lowest redshifts, focusing on the question of core-halo structure for comparison with new deep imaging and spectroscopic data for the Milky Way and other local group dwarfs.

The structure of this paper is as follows. In sections \ref{“WDM”-model} and \ref{dynamic-“WDM”} we describe the radial structure and internal dynamics of the $\psi$DM halo for comparison with the data. Next, in section \ref{sim-setup} we present the simulation setup. In section \ref{results} we focus on presenting all the results of the work, with particular emphasis on the main finding related to the detection of the core-halo structure. Finally, in section \ref{conclu} we discuss our conclusions regarding the nondetection of the core-halo structure for CDM and its implications. The boson masses used in this work are in strong tension with current constraints in the literature (for a discussion of these constraints, see section 6).

\section{The wave/fuzzy dark matter halo}
\label{“WDM”-model}
Ultralight bosons, such as ALPs (axion-like particles), have been considered as an ingredient in describing the $\psi$DM \citep{Widrow:1993,Hu:2000}. In the absence of any self-interactions, the boson mass is the only free parameter that describes the DM. For a sufficiently light boson mass, the de-Broglie wavelength exceeds the mean free path, which  leads the ALP to satisfy the ground state condition for a Bose-Einstein condensation. The governing equation for $\psi$DM is the
Schr\"{o}dinger-Poisson (SP) equation, which in comoving coordinates reads as:
\begin{align}
& \biggl[i\frac{\partial}{\partial \tau} + \frac{\nabla^2}{2} - aV\biggr]\psi=0\,,\\
& \nabla^2 V =4\pi(|\psi|^2-1)\,.
\end{align}
where $\psi$ is the wave function (see \cite{Schive:2014} for the explanation of the interpretation of the wave function, originally introduced by \cite{Widrow:1993}, to describe the mean field behavior of the wave function in this context), $V$ is the gravitation potential and $a$=(1/(1+z)) is the cosmological scale factor, where z is the redshift.The comoving length $\bm{x}$ is normalized to ($\frac{3}{8\pi} H_0^2 \Omega_{m0})^{-1/4}$ $(m_\psi/\hbar)^{1/2}$, the time normalized to $d\tau=(\frac{3}{8\pi} H_0^2 \Omega_{m0})^{1/2} a^{-2}dt$ and the wave function $\psi$ normalized to $(\rho_{m0}/m_{\psi})^{1/2}$, where $H_0$ is the present Hubble parameter,$\Omega_{m0}$ is the matter density parameter and $\rho_{m0}$ the background mass density, \citep{Schive:20142}. Finally, the physical interpretation of the wave function in $\psi$DM is that the density is given by $\rho = |\psi|^2$.

Recently, it has become possible to perform high dynamic range cosmological simulations that solve the above equations \citep{Schive:2014, Schwabe:2016, Mocz:2017, May:2021}, where GPU computing has allowed an efficient adaptive mesh refinement \citep{Schive:2014}. These evolve to produce large-scale structures indistinguishable from CDM, but with virialized halos characterized by a solitonic core in the ground state that naturally explains the dark matter-dominated cores of dSph galaxies, previously analyzed in many dSph and UFD of the Local Group \citep{Pozo:2020,Pozo:2023,Pozo:2021,Pozo:2022,Schive:20142} . Another important feature arising from the simulations is that the central soliton is surrounded by an extended halo with a granular texture on the de-Broglie scale, due to interference of excited states, but which when azimuthally averaged follows the Navarro-Frenk-White (NFW) density profile \citep{Navarro:1996,Woo:2009,Schive:2014,Schive:20142}.

The fitting formula for the density profile of the solitonic core in a $\psi$DM halo  is obtained from cosmological simulations \citep{Schive:2014,Schive:20142} as
\begin{equation}\label{eq:sol_density}
\rho_c(r) \sim \frac{1.9~a^{-1}(m_\psi/10^{-23}~{\rm eV})^{-2}(r_c/{\rm kpc})^{-4}}{[1+9.1\times10^{-2}(r/r_c)^2]^8}~M_\odot {\rm pc}^{-3},
\end{equation}
where $m_\psi$ is the boson mass
and $r_c$ is the solitonic core radius. The latter scales with both the halo mass and the boson mass, obeying the following scaling relation which was derived from the simulations (Eq. 7 \citep{Schive:20142})
\begin{equation}\label{eq:sol_radius}
r_c=1.6\biggl(\frac{10^{-22}}{m_\psi}  eV \biggr)a^{1/2}
\biggl(\frac{\zeta(z)}{\zeta(0)}\biggr)^{-1/6}
\biggl(\frac{M_h}{10^9M_\odot}\biggr)^{-1/3} {\rm kpc},
\end{equation}
where $M_h$ is the halo mass and $\zeta(z) \equiv \frac{18 \pi^2 + 82 (\Omega_m(z) - 1) - 39 (\Omega_m(z) - 1)^2}{\Omega_m(z)} \sim 350 \, (180) \text{ at } z = 0 \, (z \geq 1)$. $\Omega_m(z)$ is defined as the matter density parameter as a function of redshift, normalized to the critical density at that time. Beyond the soliton, at radii larger than a transition radius ($r_t$), the simulations also reveal that the halo roughly resembles NFW in shape, presumably reflecting the nonrelativistic nature of condensates beyond the de Broglie scale, and therefore the total density profile can be written as
\begin{equation}\label{eq:dm_density}
\rho_{DM}(r) =
\begin{cases} 
\rho_c(r)  & \text{if \quad}  r< r_t, \\
\frac{\rho_0}{\frac{r}{r_s}\bigl(1+\frac{r}{r_s}\bigr)^2} & \text{otherwise},
\end{cases}
\end{equation}

where $\rho_0$ is chosen such that the inner solitonic profile matches the outer NFW-like profile at approximately $\simeq r_t$,
and $r_s$ is the scale radius. In detail, the scale radius of the solitonic solution, which represents the ground state of the Schr\"{o}dinger-Poisson equation, is related to the size of the halo through the uncertainty principle.

\begin{figure*}
\sidecaption
\includegraphics[height=17cm]{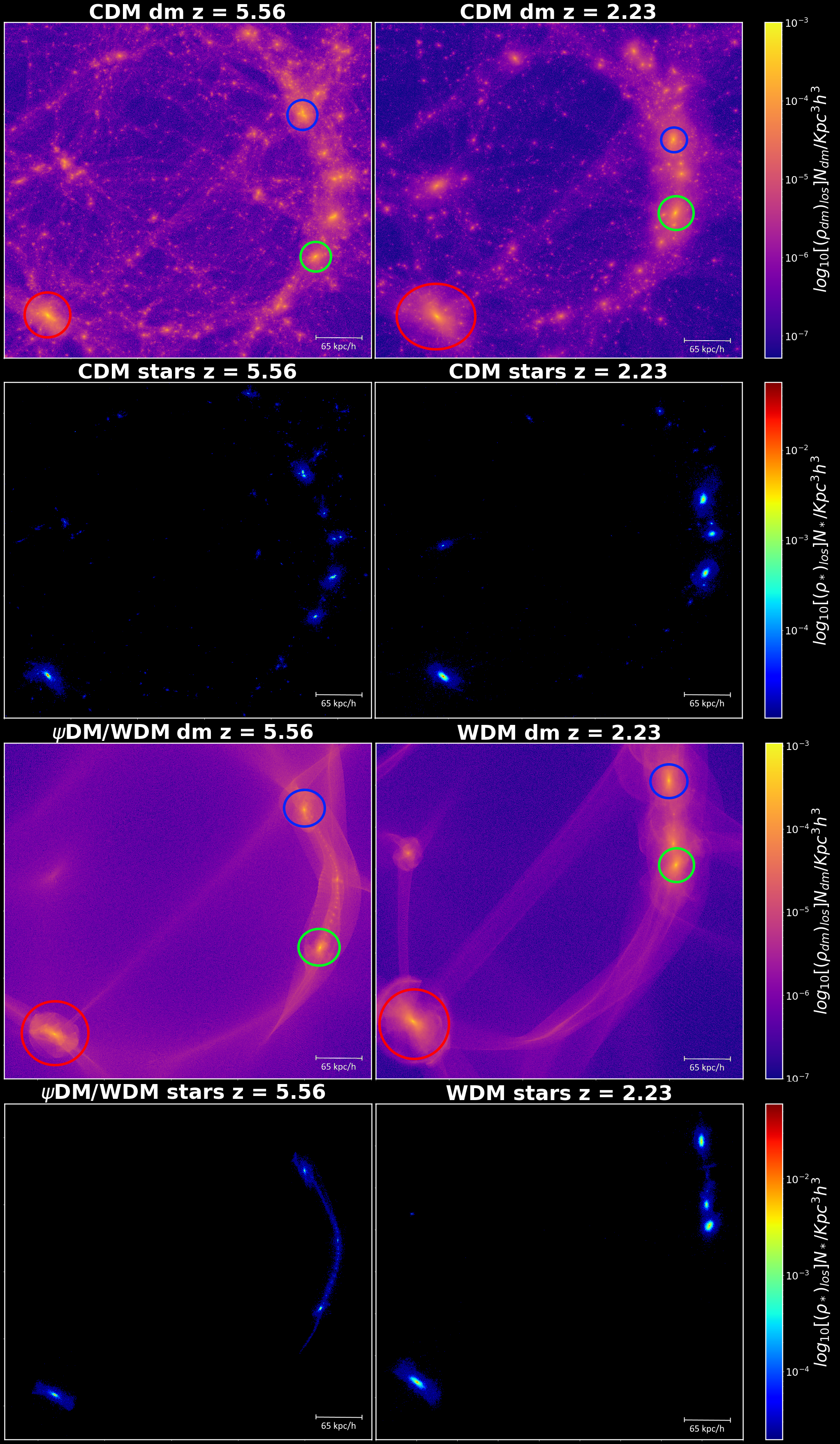}
\caption{Logarithmic (comoving) projected densities along the line of sight from the study by \citet{Mocz:2019}. The projection is performed along the entire line-of-sight dimension of the simulation box. The first row shows the profiles for DM in the context of CDM, and the second row displays the corresponding profiles for stars. The third and fourth rows illustrate the profiles for DM and stars in the context of $\psi$DM--“WDM”. In each row, the left panel represents the density profile at a redshift of $z=5.56$, while the right panel shows it at $z=2.23$. “WDM” and $\psi$DM exhibit remarkably similar evolutionary trends across different redshifts, suggesting their resemblance in terms of the development of large-scale structures, in contrast to CDM \citep{Mocz:2019,Mocz:2020}. We have added colored circles to identify each galaxy throughout this work:  Galaxy 1 (G1) marked in red, Galaxy 2 (G2) in green, and Galaxy 3 (G3) in blue. }\label{“WDM”vsWave} 
\end{figure*}

\begin{figure*}
	\centering
	\includegraphics[width=1\textwidth,height=20cm]{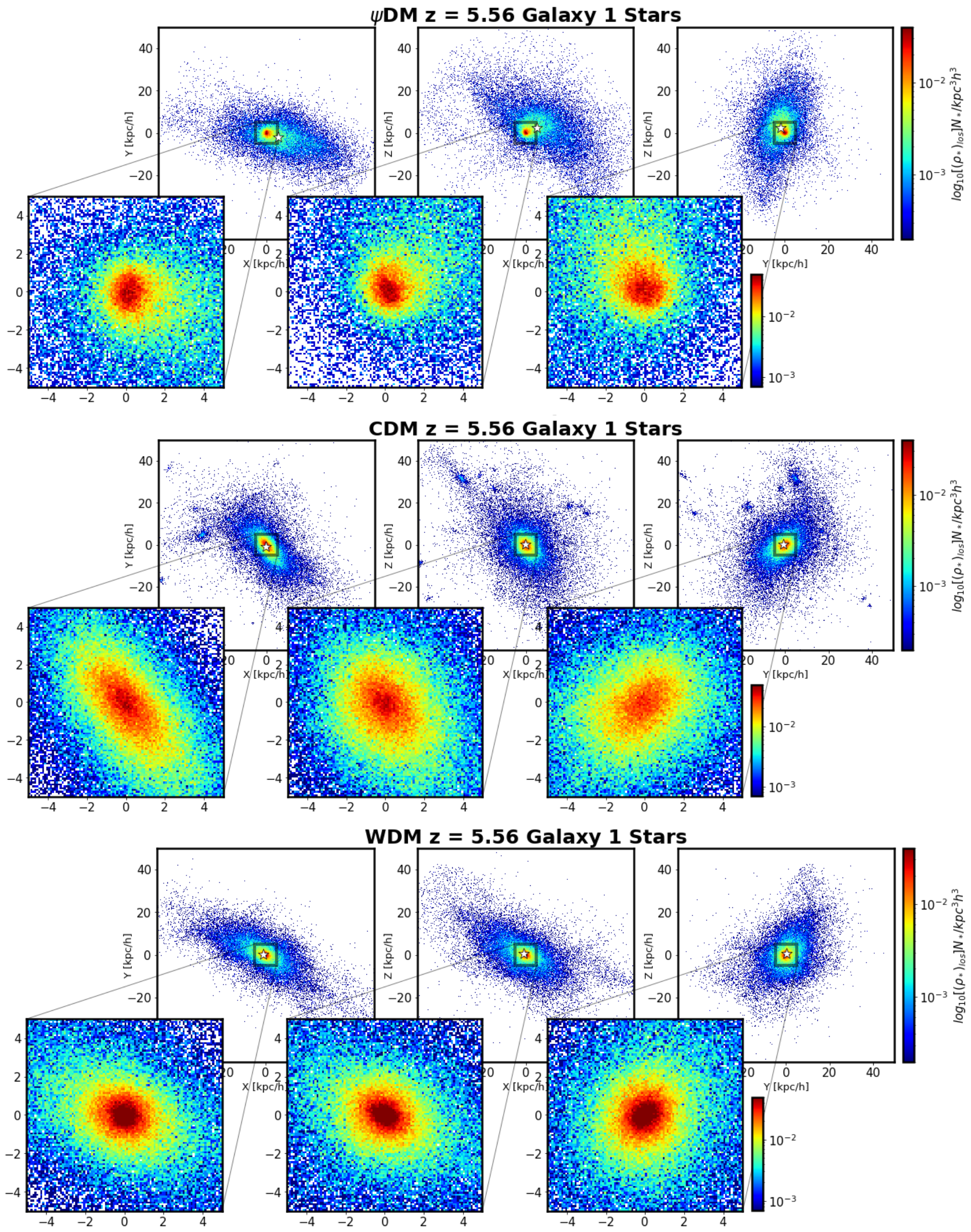}
	\caption{Logarithm of the projected number density of stellar particles in each simulated Galaxy. The row headers indicate the halo number, the DM model, and the redshift of the data. Additionally, we   zoom in on the central regions of the Galaxies for a closer view. Notably, the motion of the soliton is evident, resulting in an offset (compared to the centroid of the full 2D stellar distribution, marked by the white star) and asymmetry (exclusively addressing the observed comet tail in the core of the Galaxy) in the distribution of stellar particles, a contrast to the symmetry observed in CDM and “WDM” scenarios.  The data is represented in comoving units.}\label{starsmorpho}
\end{figure*}

\section{Dynamical model of galaxies in wave/fuzzy dark matter halo}
\label{dynamic-“WDM”}
Classical dSph galaxies are known to be dominated by DM with stars being treated as the tracer particles \citep{Gregory:2019,McConnachie:2006,McConnachie:20062,Kang:2019} moving along the gravitational potential generated by the DM halo. In this context, the corresponding velocity dispersion profile can be estimated by solving the spherically symmetric Jeans equation:
\begin{equation}\label{eq:sol_Jeans}
\frac{d(\rho_*(r)\sigma_r^2(r))}{dr} = -\rho_*(r)\frac{GM_{DM}(r)}{r^2}-2\beta\frac{\rho_*(r)\sigma_r^2(r)}{r}.
\end{equation}
Here $M_{DM}(r)$ is the DM halo mass obtained by integrating the spherically symmetric density profile in Eq. \eqref{eq:dm_density}, $\sigma_r(r)$ describes the
radial velocity dispersion, $\beta$ is the anisotropy parameter (see \citep{Binney:2008}, Equation (4.61)), and $\rho_*(r)$ is the stellar density profile as given by
\begin{equation}\label{eq:stellar_density}
\rho_{*}(r) =
\begin{cases} 
\rho_{1*}(r)  & \text{if \quad}  r< r_t, \\
\frac{\rho_{02*}}{\frac{r}{r_{s*}}\bigl(1+\frac{r}{r_{s*}}\bigr)^2} & \text{otherwise,}
\end{cases}
\end{equation}
where
\begin{equation}\label{eq:stellar_1}
\rho_{1*}(r) = \frac{\rho_{0*}}{[1+9.1\times10^{-2}(r/r_c)^2]^8}.
\end{equation}
Here, $r_{s*}$ is the 3D scale radius of the stellar halo corresponding to $\rho_{0*}$ the central stellar density, $\rho_{02*}$ is the normalization of $\rho_{0*}$ at the transition radius and the transition radius, $r_t$, is the point where the solitonic structure ends and the halo begins at the juncture of the core and halo profiles. 

We employ Eq. \eqref{eq:stellar_density} to identify the core-halo structure in the analyzed $\psi$DM stellar profiles, given its high efficiency in fitting the observed core-halo transition in local group galaxies \citep{Pozo:2020} and as it is the only profile capable of explaining this observed transition between the two regimes—inner core and outer halo—together. This approach is used exclusively for $\psi$DM, as it is unsuitable for non-$\psi$DM models, where the presence of a soliton is not expected \citep{Pozo:2022,Pozo:2023}. For CDM--“WDM”, there is no equation that describes both regimes simultaneously, as such a structure is not expected in their DM or stellar profiles. However, for the $\psi$DM profile, we can adjust eq. \eqref{eq:dm_density}
, which defines the core-halo structure (described in Section 5.3) within the $\psi$DM model, to align with what we expect for stars, as indicated by eq. \eqref{eq:stellar_density}.

For CDM/“WDM” we   used the isothermal equation from \cite{Binney:2008} to describe the observed stellar cored profiles:
\begin{equation}\label{eq:stellar_2}
\rho(r) = \frac{\rho_0}{1 + \left( \frac{r}{r_c} \right)^2}.
\end{equation}

Here $\rho(r)$ is the mass density at a radial distance r from the center of the galaxy. $\rho(r)$  represents the central density, or the density at r = 0. $r_c$ is the core radius, which defines the scale at which the density profile transitions from a flat core to a declining halo.

The isothermal profile is commonly used to describe galaxies with a central region where the density remains approximately constant (the core) and a halo where the density decreases with distance, following approximately $\rho \propto r^{-2}$ at large $r$. However, this decrease is incompatible with the core-halo structure defined in Section 5.3 and in eq. \eqref{eq:dm_density}, which predicts a transition point where the density should start declining according to an asymptotic profile of $r^{-3}$.

Finally, the predicted 2D stellar density profile can be projected
along the line of sight, and compared with the observations, as

\begin{equation}
\Sigma_*(R) =2\int_{R}^{\infty} \rho_*(r)(r^2-R^2)^{-1/2}rdr\,. 
\end{equation}

\section{Simulation setup}
\label{sim-setup}
Throughout this manuscript, we extensively make use of the most recent simulations of $\psi$DM of \cite{Mocz:2019, Mocz:2020}. These simulations adopt a simulation box of $L_{\mathrm{box}} = 1.7 h^{-1} \mathrm{Mpc}$ and a boson mass of $m_\psi = 2.5 \times 10^{-22} {\rm eV}$ and starts from an initial redshift of $z=127$ evolving to a final redshift of $z=2.23$ (5.56 for $\psi$DM). The initial conditions are generated using the publicly available Boltzmann code AXION-CAMB. The selection of the final redshift is driven by the need for a resolution to achieve fully converged results. The DM spatial resolution is set at $1024^3$, accompanied by a baryonic resolution of $512^3$. The cosmological parameters correspond to the Planck results.

The primary objective is to conduct a thorough comparative analysis of the stellar properties between the $\psi$DM simulations and the ones stemming from both CDM and “WDM” simulations, spanning an extended duration down to $z=2.23$. In the case of the “WDM” simulations, a particle resolution of $512^3$ was utilized, employing the identical hydrodynamical setup as the $\psi$DM simulations. It is noteworthy that while the CDM simulation lacks any form of cutoff in its power spectrum, both the $\psi$DM and “WDM” simulations adopt an initial power spectrum characterized by exponential suppression.

Baryons (gas and star formation) are included in the simulation, and couple to the dark matter gravitationally. The baryonic material experiences primordial and metal-line cooling, chemical enrichment, stochastic star formation with a density
threshold of $0.13~{\rm cm}^{-3}$, supernova feedback via kinetic winds, and instantaneous uniform reionization at $z\sim 6$ \citep{2013MNRAS.436.3031V,2014MNRAS.438.1985T,Pillepich:2017,2018MNRAS.475..676S}, 
as has been used in the Illustris and Illustris-TNG projects.
The stellar feedback model has previously been tuned to match key observables with CDM simulations \citep{2013MNRAS.436.3031V}. Whether FDM requires significant re-tuning of efficiency factors is left for future study. The subgrid model for feedback is meant to describe the effects of Type II supernovae (SNII),
and uses the local star formation rate to set the mass loading of 
stellar winds driven by the energy available through SNII \citep{2013MNRAS.436.3031V,2018MNRAS.473.4077P}. 
The simulation also keeps track of metallicty, as described in \cite{2019MNRAS.484.5587T}.
Star particles (representing a population of stars) are born with the metallicity of their surrounding interstellar gas. As star particles evolve in time, mass and metals from the aging
stellar populations are returned to the interstellar medium, taking into account SN1a and SNII supernovae and AGB stars. Deposited metals in the gas evolve through passive advection with the fluid flow.

\section{Results}
\label{results}

\subsection{First illustration of density profiles}
\label{illustration}
Before delving into the intricacies of comparing various simulation types, we provide an initial depiction of the projected density profiles for CDM, “WDM”, and $\psi$DM at two different redshifts. In Figure \ref{“WDM”vsWave}, we present the projected density profiles for both DM and stars. The first two rows illustrate CDM profiles, while the last two rows depict the corresponding profiles for “WDM”/$\psi$DM. In each row, the left(right) panel presents $z=5.56$ ($z=2.23$). \cite{Mocz:2019} showed that “WDM” and $\psi$DM exhibit remarkably similar evolutionary patterns across various redshifts, implying a resemblance in the development of large-scale structures. This stands in contrast to CDM. Consequently, in Figure \ref{“WDM”vsWave} we exclusively focus on presenting the “WDM” results.

Figure \ref{“WDM”vsWave} and \ref{starsmorpho} suggest slight discrepancies between galaxies in these three distinct DM scenarios. First, in the context of CDM, filaments rapidly collapse into compact halos, whereas in the case of $\psi$DM, these filaments retain a smoother configuration without giving rise to small galaxies. Second, in the CDM scenario, stars seem to exhibit greater concentration with irregular structures. Conversely, in the case of $\psi$DM and “WDM”, a flatter shape is evident, accompanied by galaxies possessing extended stellar halos. This imparts a visual impression that stars are positioned farther away from the center. As we consider individual stellar profiles of galaxies across distinct DM scenarios, a notable finding emerges. In the $\psi$DM scenario, the stellar profiles exhibit a flatter behavior in the inner regions, in alignment with the expected pattern of stars tracing the presence of a soliton in the core. In contrast, both “WDM” and CDM scenarios lack such a soliton, resulting in a cuspy inner profile. While the primary focus of this study lies in the comparison of evolved cosmological structures between $\psi$DM and CDM, it is reasonable to anticipate that the large-scale structures in “WDM” would bear similarities to $\psi$DM. As illustrated in \cite{Mocz:2019}, the resulting structures are remarkably akin for “WDM” and $\psi$DM at z=5.56, despite the more pronounced differences in their internal halo structures.

\begin{figure*}
	\centering
	\includegraphics[width=1\textwidth,height=5.5cm]{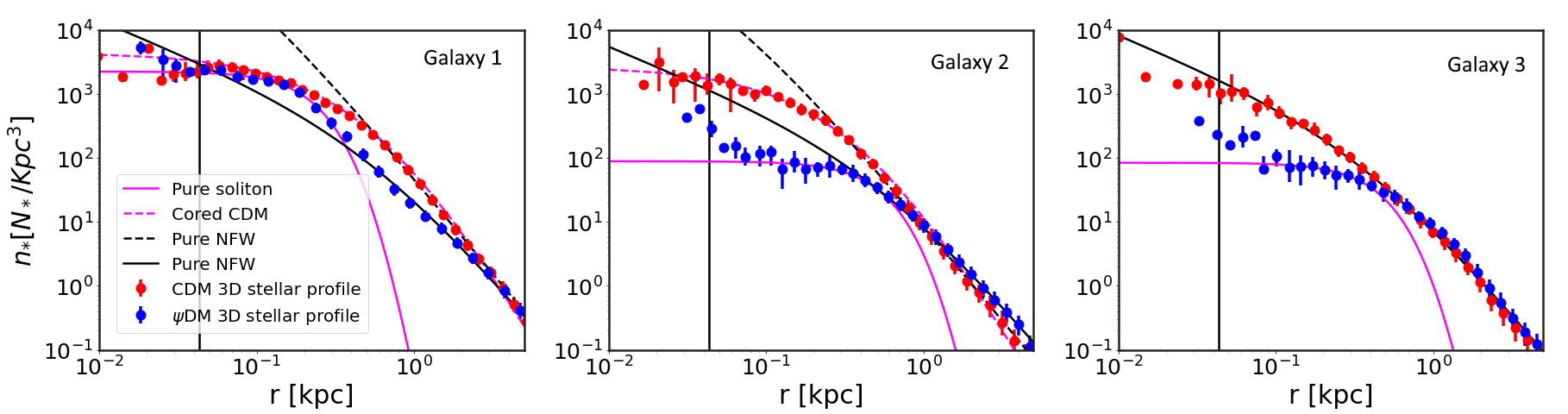}
	\caption{  $\psi$DM vs. CDM extracted stellar profiles from \citet{Mocz:2019, Mocz:2020} simulation data ($z=5.56$). This figure presents the extracted stellar profiles of the $\psi$DM and CDM Galaxies from the simulation data of \citet{Mocz:2019, Mocz:2020} at $z=5.56$. We zoom in on the inner part of the galaxy using 0.01 kpc binning. We fitted $\psi$DM profiles with a pure soliton--isothermal (magenta solid line) and NFW (black solid line) profiles to highlight the differences between the core and the outer interference pattern. The combination of these profiles is crucial for detecting a potential core-halo structure. This is evident in the case of Galaxy one, where both profiles (soliton and NFW) intersect, while still providing a good fit for the entire profile in the $\psi$DM scenario. Meanwhile, for CDM, there is no such two-regime structure necessity, where the entire profile can be described with a single NFW profile or an isothermal profile used to describe cored CDM profiles (dashed black and magenta lines, except for Galaxy 3 where the same NFW profile can be used to describe both DM models). The vertical black line represents the comoving resolution limit of the data, indicating that values smaller than this limit should be treated with caution due to their potentially unreliable nature, and  is why we did not take them into account for the fitting.}\label{figure2}
\end{figure*} 

\subsection{Stellar morphology}

In this section, we  compare the stellar distribution of each Galaxy across different DM simulations. 
\begin{itemize}
\item The main objective is to detect any potential imprint in the stellar structure of these Galaxies resulting from their respective dark matter models.

  \item Since the data is limited to high values of z, it is important to note that only the results for Galaxy 1 can be considered entirely reliable, as it is the only one to have a fully virialized halo.
\end{itemize}

Figure \ref{starsmorpho} presents the stellar morphology at three projections 
for the isolated G1 at $z=5.56$. The main panels in this plot display the stellar distribution for the entire Galaxy at a scale of approximately 55 kpc/h, while the additional subpanels provide a closer view of the central regions of the Galaxy (5 kpc/h). This allows for a detailed comparison between different stellar cores.

Figure \ref{starsmorpho} distinctly illustrates variations among three DM models, revealing that stars exhibit a more asymmetric structure in $\psi$DM compared to a slightly symmetric shape seen in “WDM” and progressively a more symmetric shape in CDM, aligning well with theoretical predictions \citep{Gao:2007, Schive:2020}. Notably, there is a significant discernible difference between the stellar distributions in “WDM” and $\psi$DM. Despite appearing similar in scale in Figure \ref{“WDM”vsWave}, stars in “WDM” are actually more flattened than in $\psi$DM. Upon closer examination in the subpanels with zoomed data, stars exhibit a spherical form in the $\psi$DM core, but a more elliptical shape toward the center for both CDM and “WDM”. The plot effectively highlights the disparities between the solitonic core ($\psi$DM) and cuspy dark matter centers (CDM)/“WDM”. Additionally, aside from having fewer stars than CDM and “WDM”, $\psi$DM has a smaller proportion of stars in the central regions (see Table \ref{tabla:collage4}). This suggests a more extended stellar profile around the soliton, in line with recent discoveries of halos around some studied dwarfs \citep{Chiti:2021,Collins:2021}. These extended halos are inherent to $\psi$DM, composed of excited states above the ground state soliton, as evidenced by the NFW form, as predicted by the $\psi$DM simulations \citep{Schive:2014}. This reflects the fundamentally nonrelativistic nature of $\psi$DM. Moving to Figure \ref{starsmorpho4}, we observe “WDM”--CDM data for z $\simeq$ 2.23. The results are comparable for both models when compared to the data from redshift z $\simeq$ 5.56, showing an expected symmetry.

The additional subpanels in Figure \ref{starsmorpho} prominently display the predicted asymmetry for $\psi$DM, a distinctive feature not observed in the symmetric distributions of stars in “WDM” and CDM models \citep{Schive:2020}. This asymmetry in $\psi$DM arises from the solitonic characteristic motion, which includes a random walk caused by interference on the de Broglie scale, where the spatial position of the center of the soliton changes with time, and the inherent oscillation 'jumping' behavior (soliton oscillations) on the scale of the de Broglie wavelength $\lambda_{B}$ or the core radius ($r_c$), owing to its standing wave nature \citep{Li:2021, Chiang:2021}. The jumps occur over a timescale of $\lambda_{B}/\sigma \simeq 10^{7}$ years \citep{Pozo:2023}  for a dSph with $r_c \simeq$ 0.3 kpc and $\sigma \simeq$ 10 km/s, values closely matching those for this Galaxy (as referred in Tables \ref{tabla:collage} and \ref{tabla:collage2}). Notably, this timescale is equivalent to the stellar crossing time of the soliton, $r_c / \sigma$. In essence, this means that stars ``sense'' the movement of the soliton and respond accordingly, resulting in the observed asymmetry \citep{Li:2021,Chiang:2021}. We are exclusively addressing the observed comet tail in the core of the Galaxy.

\subsection{Core-halo structure}
Below, we explore the presence of core-halo structures across various Galaxies. Before we proceed, it is important to define some concepts for the analysis. 

\begin{itemize}
  \item We define the core-halo structure as the characteristic resulting profile observed in all ideal $\psi$DM dark matter models, featuring an inner flat core that can be fitted with a soliton and an outer characteristic $r^{-3}$ asymptotic profile. The transition point, $r_t$, is simply the radius at which this change of regime occurs. This characteristic profile has been observed in the stellar profiles of many Local Group galaxies \citep{Pozo:2020, Pozo:2023}, motivating the search for such structure in the stellar profiles of the simulations presented in this work.

  \item To analyze the compatibility of the simulated stellar profiles with the searched core-halo structure, we fit each stellar profile using a core model for the inner region (Eq. \ref{eq:stellar_1} for $\psi$DM, and an isothermal model (\cite{Binney:2008}, Eq. \ref{eq:stellar_2}) for “WDM” and CDM, since Eq. \ref{eq:stellar_1} assumes a galaxy hosting a soliton, making it incompatible with “WDM” and CDM). For the outer halo, we use an NFW-like profile \citep{Navarro:1996}. A clear intersection between these two profiles indicates a high level of compatibility with the investigated core-halo structure.

\item This core-halo structure is theoretically anticipated in the $\psi$DM model \citep{Schive:2014} and has been observed in many local group galaxies stellar halos \citep{Pozo:2020}.

  \item Since the data is limited to high values of z, it is important to note that only the results for Galaxy 1 can be considered entirely reliable, as it is the only one to have a fully virialized halo.
\end{itemize}

\subsubsection{Signature of the core-halo structure}
We start off with comparing the stellar profiles at $z=5.56$. Figure \ref{figure2} compares the stellar density profiles of the three main Galaxies in the $\psi$DM model at $z=5.56$ with those of the CDM model. Moving from left to right, we present the first, second, and third Galaxy, respectively, as was previously pointed out in Figure \ref{“WDM”vsWave}. In this figure, a notable difference is seen between the stellar profiles in the $\psi$DM scenario and CDM. This highlights how stellar profiles can vary and evolve differently in distinct DM scenarios, suggesting they could serve as a valuable tool for analyzing DM signatures in dwarf galaxies. If we take a closer look at the inner part of the Galaxy we can make a clear test to assess the compatibility of the profiles with a core-halo structure. To this end, we plot the ideal solitonic/isothermal and NFW profiles that would correspond to the inner core and the characteristic $r^{-3}$ asymptotic profile, respectively. It is worth noting that only in the case of $\psi$DM$_{G1}$, for figure \ref{figure2}, do the NFW and solitonic/isothermal profiles intersect while still fitting the stellar profile. This is never a possibility for any CDM Galaxy, where the entire profile can be fitted by a single NFW or isothermal profiles. This result may indicate the potential presence of a primordial core-halo stellar structure. There is also a noticeable difference in the inner profile for Galaxies two and three, which aligns with the presence or absence of a soliton, resulting in a core-cuspy disagreement. Furthermore, in the case of Galaxy three, where both $\psi$DM and CDM exhibit a similar halo with the same slope, we can observe how the NFW profile effectively fits the entire CDM stellar profile. This is a consequence of the anticipated cusp, whereas a soliton is required for $\psi$DM to establish its flat inner profile. The reason why both stellar profiles appear to coincide with a flat core for Galaxy 1 in the inner part could be attributed to the widely suggested notion that baryonic feedback may lead to the development of a core, even in the cold dark matter (CDM) scenario \citep{Chan:2015, Pontzen:2012}. This is more plausible at this redshift for G1 than for G2 and G3, given its greater mass, as the inner slope of the dark matter (DM) halo profiles shows a strong mass dependence \citep{Chan:2015}. However, CDM G2 also requires a cored fit, albeit much smaller than the large flat inner core of $\psi$DM G2.  It is likely the same reason why the core-halo structure had sufficient time to form only in this Galaxy. The vertical line in the figure represents the comoving softening length, which is 0.19 kpc in this case. Below this radius, 
cautious must be taken in interpreting the data.

\begin{figure*}
	\centering
	\includegraphics[width=1\textwidth,height=5.5cm]{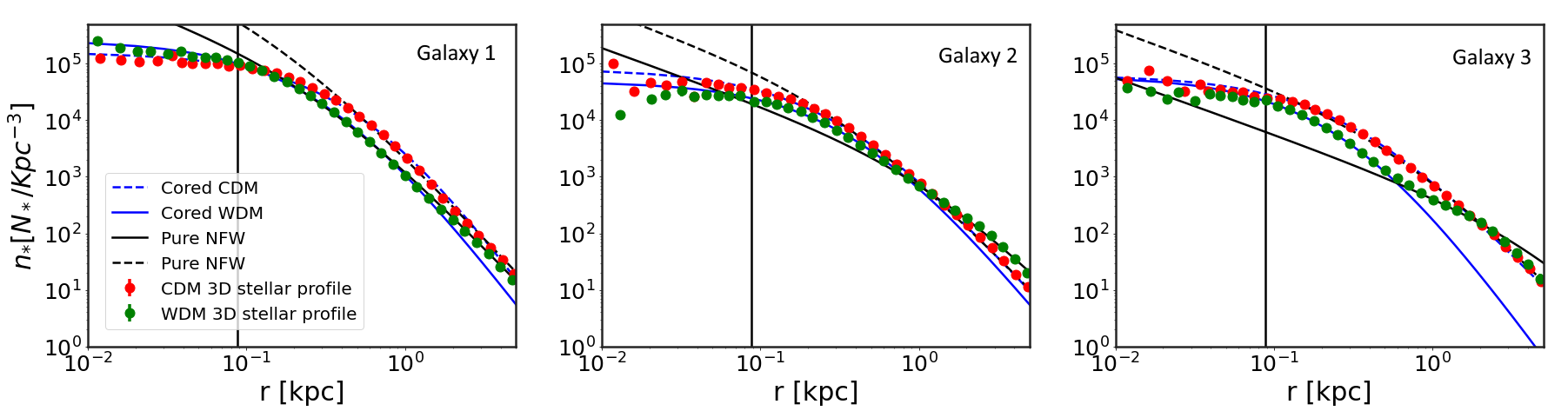}
	\caption{“WDM” vs. CDM extracted stellar profiles from \citet{Mocz:2019, Mocz:2020} simulation data ($z=2.23$). We zoom in on the interior part of the Galaxy. In this case, there is not a significant discrepancy between the two models in the inner region, indicating that CDM profiles need to be described by the presence of a core, as for “WDM”, even if it contradicts expectations for CDM--“WDM”. However, a noticeable difference becomes apparent near the theoretical $\psi$DM transition point in all three profiles, where the presence of a transition point similar to that expected for the $\psi$DM profiles is easily identifiable, particularly in the case of G3. This is particularly interesting as “WDM” does not have a dark matter core-halo structure as $\psi$DM does.  Specifically the solitonic cored profile. The vertical black line represents the comoving resolution limit of the data, indicating that values smaller than this limit should be treated with caution due to their potentially unreliable nature. This is why we did not take them into account for the fitting.}\label{CFA“WDM”223}
\end{figure*}

Next, we proceed with comparing the stellar profiles at $z=2.23$ between the “WDM” and CDM simulations, following the approach outlined in the previous section. It is worth noting that there was no available data at this redshift for $\psi$DM. Despite the significant differences between “WDM” and $\psi$DM on small scales where the core-halo structure is anticipated, their similar large-scale evolution and initial conditions could make the comparison between “WDM” and CDM valuable for future studies. The resulting extracted stellar profiles, shown in Figure \ref{CFA“WDM”223}, seem to reaffirm the same findings as the previous redshift for CDM. In CDM, there is no evidence or imprint of a core-halo structure  identified in the halos of the three simulated Galaxies. In contrast, one “WDM” halo exhibits a potentially compatible core-halo stellar profile, particularly in a primordial context, suggesting that they may still be in the process of forming for Galaxies one and two. Furthermore, a well-defined core-halo structure is evident for the third Galaxy, characterized by a notable density drop from the core to the halo, along with the distinctive change of regime point (transition point, $r_t$) marking the boundary between the inner core and the outer halo. This is observable where the pure NFW and cored profiles intersect. It is interesting to observe that for “WDM”, the core-halo structure emerges earlier in a lower mass galaxy, such as Galaxy three, in comparison to the more massive Galaxy one. This structure should be established by a more pronounced core, likely a combination of a dense soliton and potentially baryonic feedback, both more prominent in heavier galaxies. Nevertheless, it is crucial to remember that “WDM” lacks a solitonic core. The presence of a core-halo structure remains intriguing for “WDM” and that's the main reason why we only proceed to evaluate it for $\psi$DM$_{G1}$, figure \ref{CFAMCMC}.

\begin{figure*}
\centering
\includegraphics[width=1\textwidth,height=15cm]{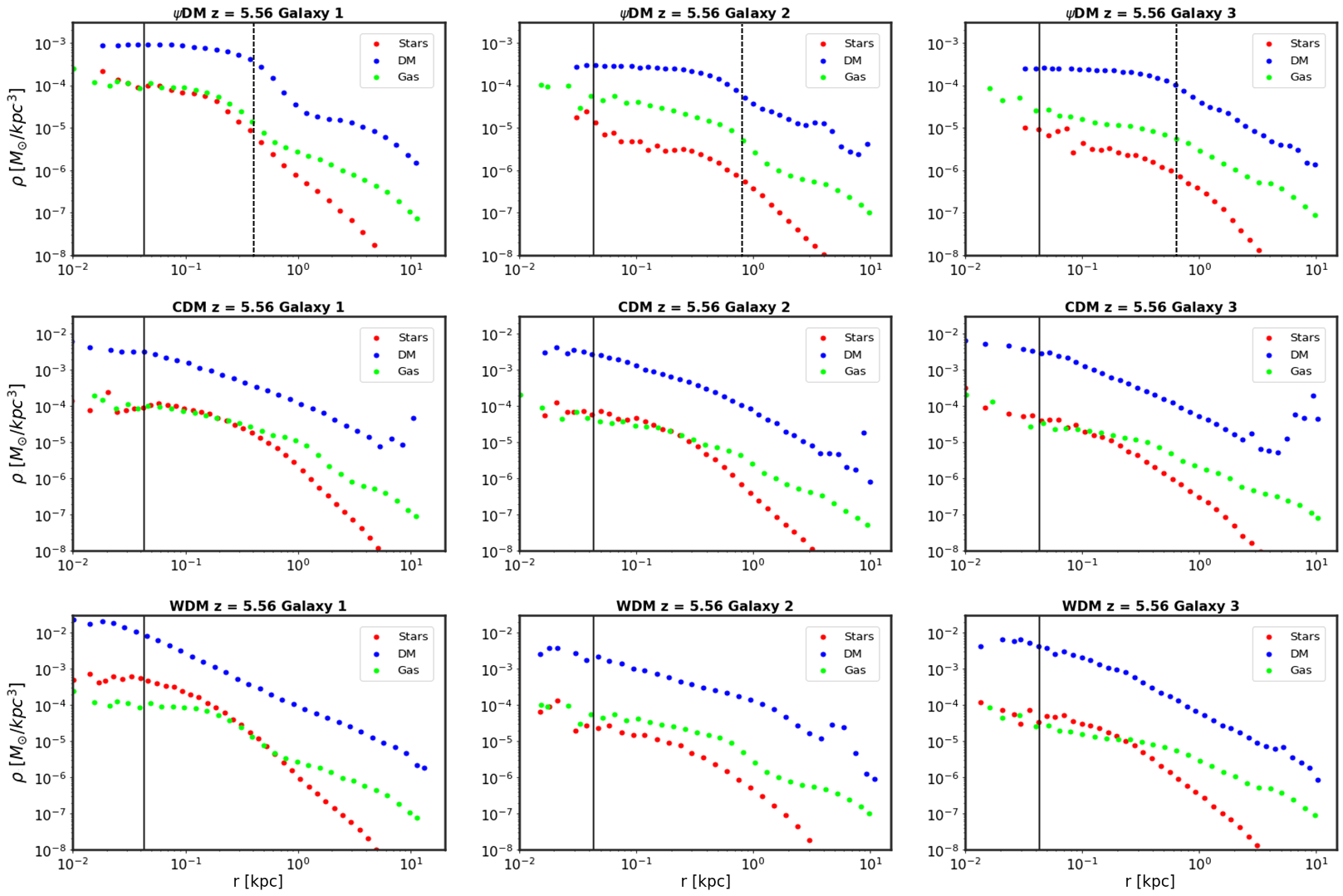}
\caption{ Representation of the stellar, dark matter, and gaseous profiles for three Galaxies from different DM scenarios: $\psi$DM, “WDM”, and CDM. The core-halo structure is visible for all three $\psi$DM dark matter profiles with a clear transition point dividing both regimes, while CDM and “WDM” exhibit the expected cuspy shapes. This difference is also visible in stellar profiles between different DM models, with bigger flat cores and more prominent falls in the $\psi$DM cases compared to the more smoothed shapes of “WDM”--CDM. For $\psi$DM G1, the stellar and gaseous profiles display a similar core radius as well as transition points compared to the DM profile. The vertical dashed black line represents the core radius of the respective solitons. The vertical solid black line represents the comoving resolution limit of the data, indicating that values smaller than this limit should be treated with caution due to their potentially unreliable nature. This is why we did not take them into account for the fitting}\label{all}
\end{figure*}

\subsubsection{Stellar, dark matter, and gas distribution}
Here we compare distinct distributions of stellar, gaseous, and dark matter profiles in all three different simulated Galaxies across different DM scenarios. The aim is to extend the former comparison made only for the stellar profiles, done in Figure \ref{figure2} and \ref{CFA“WDM”223}, to further include the DM and gaseous distributions as well. Figure \ref{all} presents these profiles at $z=5.56$. Once again, the stark contrast between the stellar and DM profiles in each case is evident, with gas profiles exhibiting a more consistent pattern across different DM models. Notably, a common feature is the presence of a core-halo transition in the DM profile from $\psi$DM and gas profiles, even if it is only visibly prominent in the case of stars in G1. In comparison with $\psi$DM, “WDM” and CDM exhibit strikingly similar cuspy DM profiles with smaller flat inner regions, while the gas profile resembles that of the $\psi$DM case. From the plot, it is inferred that in $\psi$DM model both of the stellar and gaseous profiles demonstrate similar core radii and transition points compared with the DM profile (for G1). This suggests that stars, rather than gas, may hold the key to determine the underlying DM model. Another crucial observation is the nearly identical core size and transition point in stars and gas, closely mirroring that reproduced by the DM particles (for $\psi$DM G1). This implies a potential baryonic-dark matter connection and hints at the possibility of baryonic tracking of dark matter particles \citep{Pozo:2021,Pozo:2022,Pozo:2020,Pozo:2023}. It is also important to remark that stellar profiles exhibit differences among the three DM models, with larger flat cores and more prominent declines in the $\psi$DM cases compared to the smoother shapes of “WDM”--CDM. This reveals that stellar profiles evolve and behave slightly differently depending on the DM models, reinforcing their potential utility in distinguishing between the underlying DM models.

\begin{figure*}
\centering
\includegraphics[width=1\textwidth,height=7cm]{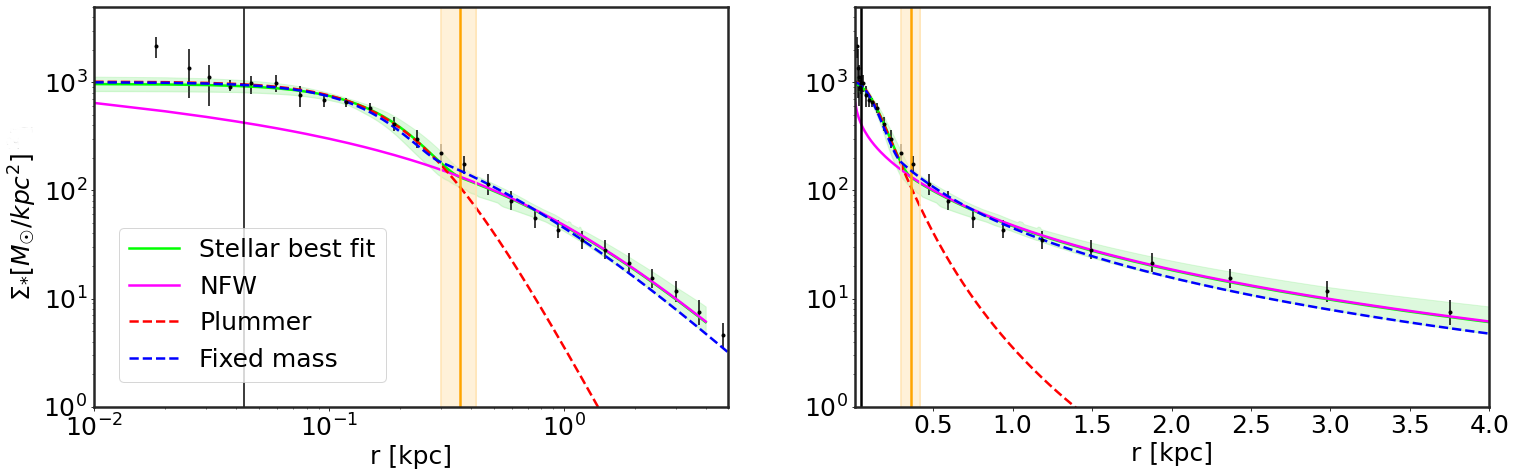}
\caption{   Isolated $\psi$DM$_{G1}$ ($z=5.56$). This figure presents the best-fit projected star count profile of the $\psi$DM isolated dwarf Galaxy, specifically G1 from Figure \ref{“WDM”vsWave}. The plot reveals an extended halo of stars, reaching approximately 4 kpc, which is most prominently displayed in the linear scale representation on the right panel. Additionally, a distinct core is evident on a scale smaller than 0.5 kpc. These characteristics closely align with the observed stellar profiles of real dwarf spheroidal galaxies in the Local Group \citep{Pozo:2023} and are consistent with the recently detected extended stellar halos in several local dwarfs \citep{Chiti:2021,Collins:2021,Torrealba:2019}. A standard Plummer profile (indicated by a red dashed curve) approximately fits the core region but falls significantly short at larger radii. Our predictions for the dSph class ($2.5 \times 10^{-22}$eV) in $\psi$DM are depicted in green (representing the 2$\sigma$ range of the posterior distribution of profiles), where the distinctive soliton profile provides an excellent fit to the observed cores and the surrounding halo of excited states. This averages azimuthally to an approximately NFW-like profile beyond the soliton radius. The accuracy of the core fit to the soliton is best visualized on a logarithmic scale in the left panel, while the right panel demonstrates the extent of the halo, including the characteristic density drop of approximately a factor of 30 predicted by $\psi$DM between the prominent core and the tenuous halo at a radius of around 0.5 kpc, indicated by the vertical orange band. The blue dashed curve represents our prediction with a fixed mass of $8.3 \times 10^{9} M_{\odot}$, which exactly matches the calculated mass by \citet{Mocz:2020} for this Galaxy at this redshift. The vertical black line represents the comoving resolution limit of the data, indicating that values smaller than this limit should be treated with caution due to their potentially unreliable nature. This is why we did not take them into account for the fitting}\label{CFAMCMC}
\end{figure*}

\begin{table*}
\caption{Profile parameters for dwarfs associated with the Milky Way and Andromeda.}
	\centering
\begin{tabular}{|c|c|c|c|c|}
\hline
Combinations & $r_c$  & $r_{t}$& $z$ &   $m_{\psi}$  \\
& (kpc)  & (kpc) & & $10^{-22}$eV\\
\hline
$\psi$DM$_{G1}$  &$0.155^{+0.021}_{-0.017}$ &$0.36^{+0.064}_{-0.056}$ &5.56 &$2.5$\\
\hline
dSph$_{\rm Both}$  &$0.21^{+0.003}_{-0.003}$ &$0.71^{+0.021}_{-0.021}$ &0 &$1.85^{+0.66}_{-0.58}$\\
\hline
dSph$_{\rm Milky\,Way}$  & $0.22^{+0.003}_{-0.003}$&$0.75^{+0.022}_{-0.023}$ &0 &$1.85^{+0.66}_{-0.58}$\\
\hline
dSph$_{\rm Andromeda}$  &$0.26^{+0.007}_{-0.006}$ &$0.82^{+0.032}_{-0.028}$ &0 &$1.86^{+0.45}_{-0.53}$\\
\hline
Draco  &$0.17^{+0.01}_{-0.01}$ &$0.56^{+0.02}_{-0.02}$ &0 &-\\
\hline
Leo II  &$0.17^{+0.01}_{-0.01}$ &$0.66^{+0.02}_{-0.01}$ &0 &-\\
\hline
Sculptor  &$0.21^{+0.01}_{-0.01}$ &$0.72^{+0.07}_{-0.07}$ &0 &-\\
\hline
And IX  &$0.22^{+0.02}_{-0.02}$ &$0.70^{+0.08}_{-0.08}$ &0 &-\\
\hline
And XV  &$0.19^{+0.02}_{-0.02}$ &$0.65^{+0.08}_{-0.09}$ &0 &-\\
\hline
\hline
\hline
\end{tabular}
\tablefoot{Column 1: Dwarf Class or Dwarf individual name, Column 2: Core radius  $r_c$,  Column 3: Core-Halo transition radius $r_t$, Column 4: Redshift $z$, Column 5: Boson mass $m_{\psi}$. Even though the core-halo structure also seems to emerge in the cosmological evolution of “WDM” G3, we have decided not to introduce it in the analysis due to discrepancies with theoretical expectations, where two different physical regimes are not expected, as in $\psi$DM.} 

\label{tabla:collage}
\end{table*}

\begin{figure*}
\centering
\includegraphics[width=1\textwidth,height=7cm]{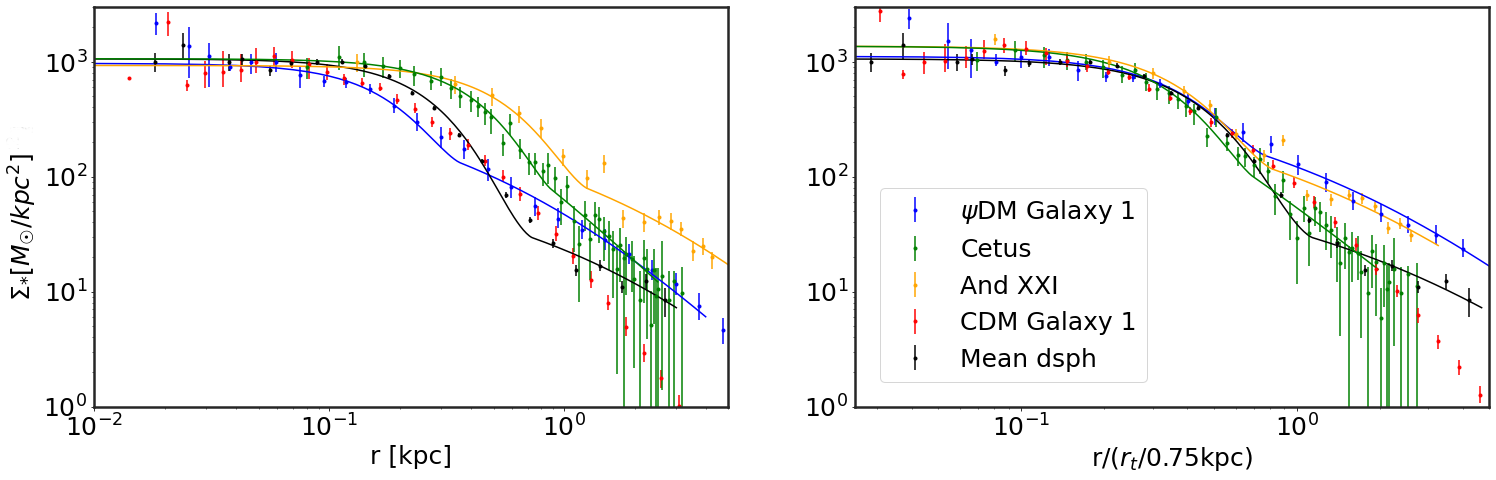}
\caption{ Comparison of stellar profiles from various galaxies. Specifically, it compares the stellar profile of the $\psi$DM and CDM isolated Galaxy (G1 from Figure \ref{“WDM”vsWave}), the mean profile of all dSphs in the Local Group \citep{Pozo:2023}, and the projected density profiles of two galaxies from Andromeda and the Milky Way that have the most similar density gap between the core and the transition point when compared to G1.  In the left panel, the four profiles are normalized based on their peak density values.  In the right panel, they are additionally normalized with respect to the transition point. Notably, the extracted profile from the data (represented by the blue line) exhibits a shape very similar to the extracted profiles of the real observable galaxies. Furthermore, this data's profile appears to coincide extremely well with the profile of Andromeda XXI. We note that even the CDM profile of Galaxy 1 aligns well with the inner parts of other galaxies. However, the absence of any transition point in the CDM profile of Galaxy 1 makes it incompatible with observed cases, in contrast to the $\psi$DM profile.}\label{CFAcompa}
\end{figure*}

\begin{table*}
\caption{Summary of virial halo masses and radii at $z=5.56$ and $z=2.23$ under different cosmologies.}
	\centering
\begin{tabular}{|c|c|c|c|c|}
\hline
Halos& $M_{200} (z=5.56)$  & $M_{200} (z=2.23)$& $r_{200} (z=5.56)$  & $r_{200} (z=2.23)$ \\
& $10^{9}$ $M_{\odot}$  & $10^{9}$$M_{\odot} $ &kpc & kpc\\
\hline
$\psi$DM$_{G1}$ &8.3 &- &52&-\\
\hline
$\psi$DM$_{G2}$ &5.5 &- &45&- \\
\hline
$\psi$DM$_{G3}$  &5.9 &-&46& -\\
\hline
CDM$_{G1}$  & 17&39.5&65& 83\\
\hline
CDM$_{G2}$ &4.5 &26.8&42&73 \\
\hline
CDM$_{G3}$  &8.3 &26.8 &52&73 \\
\hline
“WDM”$_{G1}$  &13 &35.5 &60&80 \\
\hline
“WDM”$_{G2}$  &7.3 &38.2&49& 82 \\
\hline
“WDM”$_{G3}$ &6.9 &24.8&48& 70 \\
\hline
\hline
\hline
\end{tabular}
\tablefoot{ $M_{200}$ is the total mass of each halo at mean density over the critical density of $\Delta$c = 200, where $\Delta$c is the overdensity constant whose exact value depends on the cosmology and it is typically assumed to be 200. Values in comoving units. }
\label{tabla:collage2}
\end{table*}

\subsubsection{Estimation of the core-halo parameters}
We now proceed with MCMC-based $\psi$DM stellar profile fits (Eq. \eqref{eq:stellar_density},) to the $\psi$DM$_{G1}$, as shown in Figure \ref{CFAMCMC} ( Corner plot Figure \ref{cornerCFA}). The core radius ($r_c$), as defined in Eq. \eqref{eq:sol_radius}, plays a crucial role in characterizing the shape of the solitonic core profile. For the halo, we utilize a NFW profile with a scale radius ($r_{s*}$) and the normalization factor ($\rho_{0*}$). The transition radius ($r_t$), defining the point of density transition between the soliton and NFW profiles, is the only other free parameter. We vary this within a prior range determined by $\psi$DM simulations, specifically $2$ to $4$ times $r_c$ \citep{Schive:2014,Schive:20142}. The boson mass stands as the sole fixed parameter, maintaining a value of $2.5\times 10^{-22}$eV, consistent with that used for the simulations \citep{Mocz:2020}. The best-fit yields a resulting profile with a total mass of $\left(5.8^{+0.24}_{-0.40}\right) \times 10^9 M{\odot}$, aligning well with the computed mass from \citet{Mocz:2020} of $8.3 \times 10^9 M_{\odot}$ for this Galaxy at this redshift. In Figure \ref{CFAMCMC} the blue dashed line presents the resulting profile when we conduct the analysis with the fixed total mass of $8.3 \times 10^9 M_{\odot}$, which corresponds to the total mass of that specific simulated halo (indicated by a red circle in Figure \ref{“WDM”vsWave}) at this redshift. It is worth noting that while it may not match the accuracy of the green profile with a freely varying mass, it still demonstrates excellent agreement. Furthermore, the resulting values for $r_c \left(=0.15^{+0.021}_{-0.017}\right)$ kpc and $r_t \left(=0.36^{+0.064}_{-0.056}\right)$ kpc, fall within the expected range for such dSphs, and are very close to the directly inferred soliton radius  from the simulation ($r_c \simeq 0.14$ kpc \citep{Pozo:2020,Pozo:2023}). This concordance between the suggested soliton size based on stellar behavior and the actual size of the simulated DM halo soliton may hold significant implications for our understanding of how stellar behavior is influenced by dark matter.

The resulting values for core radius ($r_c$) as well as the transition point ($r_t$) are summarized in Table \ref{tabla:collage}, alongside other pertinent variables. It is noteworthy that these parameter values fall within the expected range when compared to those extracted from observed profiles of Local Group classical dSphs. This is particularly striking considering they were analyzed at different redshifts and are still in the process of the formation. Table \ref{tabla:collage} also provides the parameter values for the two extracted galaxies and the observed dSphs in the Local Group \citep{Pozo:2023}. The core radius values are in excellent agreement with the mean values obtained for such dSphs in the Local Group.

The isolated Galaxy (G1) from the $\psi$DM simulation data ($\psi DM_{1}$, Table \ref{tabla:collage}) exhibits a core radius akin to that of two Milky Way satellite galaxies, Leo II and Sculptor. Conversely, the third Galaxy from the “WDM” simulation data (third panel of Figure \ref{CFA“WDM”223}) seems to align well with the observed cores of Sculptor and And IX. Nonetheless, there are slight disparities in the transition point, which is reasonable given the difference in redshift and the fact that these simulated Galaxies are still in the process of formation. Notably, despite the significant redshift difference (almost 6 between the observed galaxies and the observable Local Group dwarfs), the $r_c$ values remain similar. This suggests that for a $\psi$DM origin, the soliton is likely the first structure to form in this core-halo structure.

We have included the virial masses of all these three simulated Galaxies at both redshifts in Table \ref{tabla:collage2} to facilitate a clearer comparison of their evolution across different DM models. Additionally, this table serves as a useful tool for comparing the masses of G1 (isolated) with those in G2 and G3 within the filamentary structure, as depicted in Figure \ref{“WDM”vsWave}.

\subsubsection{Comparison with observations}
We proceed by comparing the extracted $\psi$DM and CDM  G1 profiles with two known dwarf galaxies, Cetus and Andromeda XXI, that orbit the Milky Way and Andromeda, respectively, and exhibit the most similar density gaps (see Figure \ref{CFAcompa}). In the left panel, the profiles are normalized based on their stellar peak density, while the right panel is additionally normalized by their respective transition radii. The black profiles represent the mean stellar profile of all Local Group dwarf spheroidal galaxies \citep{Pozo:2023}. The right panel demonstrates that the extracted $\psi$DM stellar profile (blue profile) aligns well with a real observed stellar profile, Andromeda XXI (orange profile). This indicates that the extracted profile can reproduce observed dSph structures, in contrast to the CDM profile, here represented in red.

Furthermore, this panel reveals that the inner portions of the profiles are in good agreement, suggesting that they are consistent with a solitonic core. However, it is crucial to note that Andromeda XXI and the extracted profile are at different redshifts, 5.5 and 0 respectively. This indicates that the extracted profile still has much more time to evolve, and its final core-halo structure at redshift 0 could substantially change over time.

\begin{figure*}
	\centering
	\includegraphics[width=1\textwidth,height=16cm]{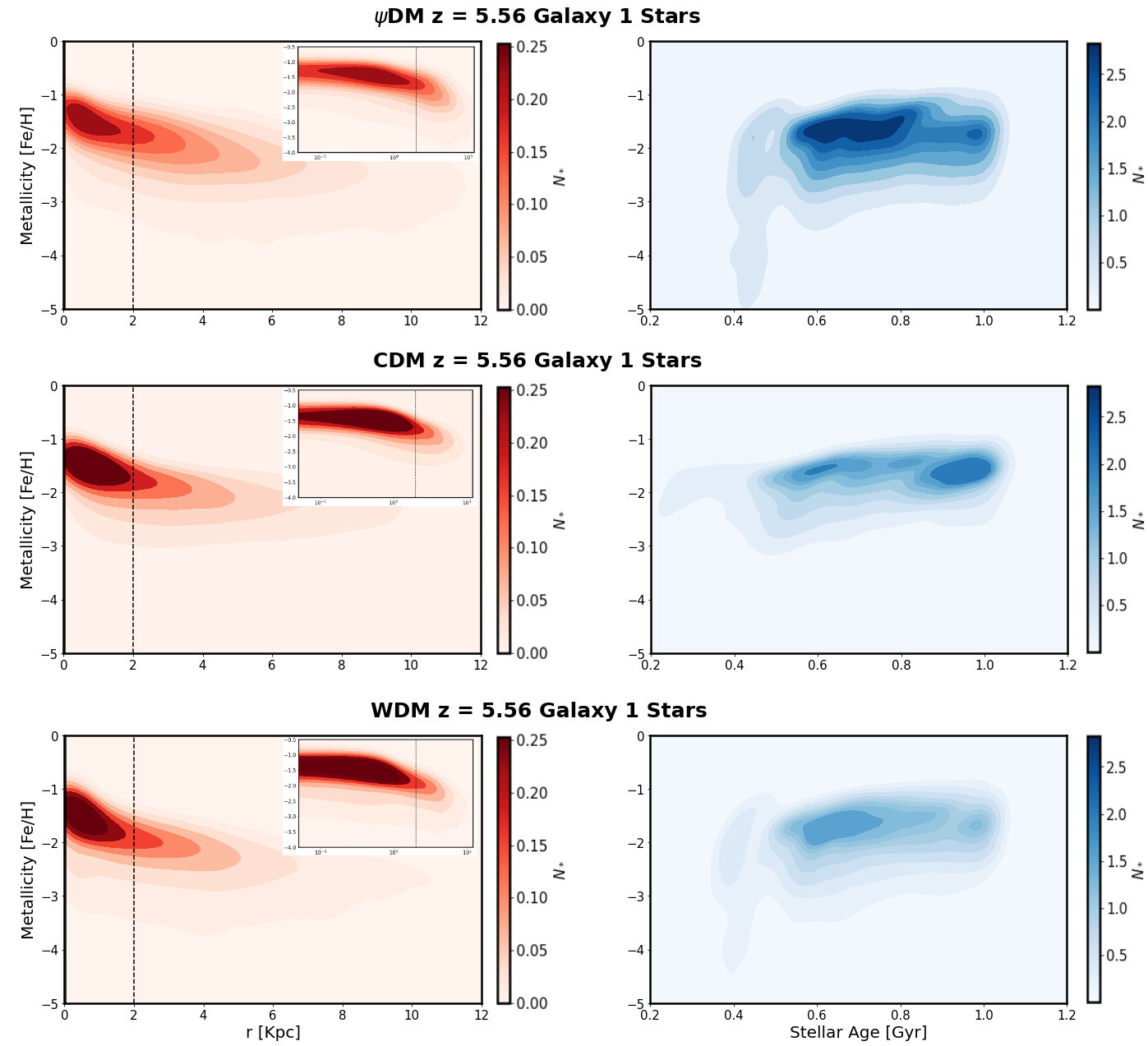}
	\caption{Left panel: 2D correlation of the stellar distance vs. stellar metallicity. It is evident that metal-rich stars tend to be located closer to the galactic center. This pattern holds across all DM models, with a slight offset observed for $\psi$DM. The small panels offer a logarithmic scale representation of the figure, providing additional insights. The vertical black line indicates the comoving limit, while the dashed line represents the ideal size of a core-halo structure.  Right panel: 2D correlation of stellar age vs. metallicity. The Spearman correlation is negative, indicating that metal-poor stars are generally older than their metal-rich counterparts \citep{Emami:2022}. Notably, “WDM” and $\psi$DM exhibit a similar stellar history, while CDM displays a more discrete profile. These visualizations offer valuable insights into the relationships between stellar properties, including distance, metallicity, and age, within the context of different DM models. }\label{age}
\end{figure*}

\begin{figure*}
	\centering
	\includegraphics[width=1\textwidth,height=16cm]{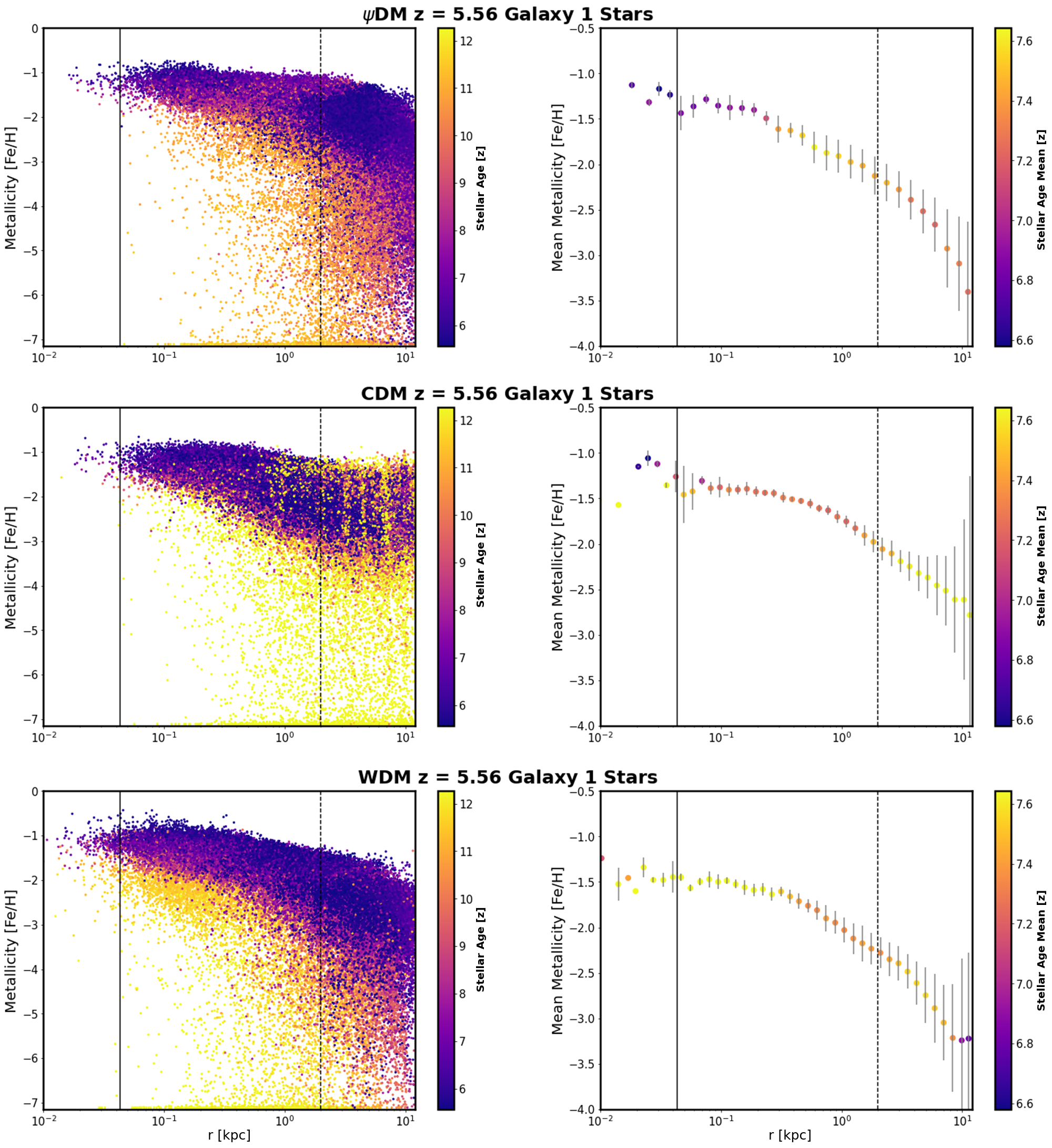}
	\caption{Left panel: 2D correlation between stellar distance from the galactic center and stellar metallicity. It is evident that metal-rich stars tend to be located closer to the galactic center, with a slight offset for $\psi$DM. The data points are color-coded based on their stellar age, providing further insights into the stellar population.  Right panel: 2D correlation of stellar spatial distribution vs. mean metallicity. The Spearman correlation is negative, indicating that metal-poor stars are generally older than their metal-rich counterparts. The data points are color-coded based on their mean stellar age, offering a comprehensive view of the age-metallicity relationship. The vertical black line marks the comoving limit, while the dashed line indicates the expected size of an ideal core-halo structure. These visualizations provide valuable insights into the relationships between stellar properties, metallicity, and age in different galactic environments and DM scenarios.}\label{starsmorpho5}
\end{figure*}

\subsection{Stellar age vs. metallicity}
Metallicity plays a crucial role in unraveling the galaxy history. Here, we explore the relationship between the stellar metallicity and age. The corresponding data is displayed in the right columns of Figures \ref{age} and \ref{age2}. These figures illustrate the [Fe/H] versus stellar age profiles for G1 at the $z=5.56$ and for Galaxies 1 and 3 at the $z=2.23$. We chose to focus on these specific Galaxies to emphasize the most pertinent data associated with instances where the potential presence of a core-halo structure has been identified.

We observe that  $\psi$DM and “WDM” have pretty similar profiles with the metal-poor concentrations at similar ages, relaying a similar stellar history. On the other hand, CDM displays a more pronounced breaks(more discrete) age vs. [Fe/H] profile, suggesting that its associated stars may have diverse origins. This disparity could arise from a combination of a few different factors. In CDM, the stellar cycle is more extended, with the first stars forming around z$\simeq$30 ( stellar age $\simeq$0.2) and continuing up to 1.2 billion years (as seen in Figure \ref{age}). In contrast, in “WDM”/$\psi$DM, star formation begins around z$\simeq$13 ( stellar age $\simeq$0.5) \citep{Mocz:2019,Mocz:2020}. Additionally, DM structures form earlier in CDM compared to “WDM”/$\psi$DM, a consequence of the small-scale cutoff in the power spectrum.

Continuing with this idea, all three Galaxies exhibit a peak of star formation around 0.6 billion years and 1.0 billion years, with the former being the predominant one for “WDM”/$\psi$DM and the latter for CDM. Additionally, CDM is the only model with an additional small peak at 0.85 billion years. As the current data does not cover more than 1 billion years, we cannot confirm the expected negative correlation between stellar age and metallicity. Similar data and predictions are shown in Figure \ref{age2} at $z=2.23$.

The observations in Figures \ref{age} and \ref{age2} suggest that metal-rich stars tend to be more concentrated in the central regions of the Galaxies, while metal-poor stars are located farther out, which is consistent with previous studies \citep{Dietz:2020,Santucci:2020}. Additionally, the majority of stars are concentrated within distances compatible with the core-halo structure (r < 2 kpc, marked with a vertical dashed line). Of particular interest is that the Galaxies where a compatible core-halo structure has been identified ($\psi$DM$_{G1}$ at $z=5.56$ and “WDM”$_{G3}$ at $z=2.23$) exhibit a radially offset region of stellar-density enhancement, which is especially pronounced in “WDM”$_{G3}$ at $z=2.23$ (as seen in Figure \ref{age2}). Notably, “WDM”$_{G3}$ is the only case where this concentration of stars extends beyond the theoretical limit of the core-halo structure (r > 2 kpc). This peculiarity may be attributed to the filamentary environment observed for this Galaxy, an environment that cannot be reproduced in the CDM model (as shown in Figure \ref{“WDM”vsWave}). In a future work we will study this in more detail. In summary, galaxies with a higher concentration of stars extending along their halos, rather than being centrally concentrated, seem to be more likely to develop the core-halo structure.

Figures \ref{starsmorpho5} and \ref{starsmorpho6} investigate the correlation between the mean stellar metallicity and the stellar radii for Galaxies from different DM models at both redshifts. 
The mean stellar age is depicted in the color bars.

The analysis of the correlation between stellar age and the spatial distribution of metallicity (as shown in Figures \ref{starsmorpho5} and \ref{starsmorpho6}) reveals distinct differences among the three DM models. One notable distinction is that in CDM and “WDM” stars are relatively more metal-rich ([Fe/H] $\approx$ -1.0) compared to $\psi$DM ([Fe/H] $\approx$ -1.25). Furthermore, CDM and “WDM” exhibit a smoother, continuous decline in metallicity from the center to the outer halo, whereas $\psi$DM displays a more discrete profile, with noticeable drops in metallicity around 0.3 kpc and 0.5 kpc (as observed in the right panel of the first row in Figure \ref{starsmorpho5}). Regarding the spatial distribution of stellar age, “WDM” exhibits older stars concentrated in the center, while the mean age decreases toward the outer parts of the halo. This trend is opposite to what is observed for CDM and $\psi$DM. However, it is important to note that for $\psi$DM, the mean age is lower compared to the mean age observed for CDM, which aligns with the predicted stellar history between both models \citep{Mocz:2019,Mocz:2020}. Tables \ref{tabla:collage3} and \ref{tabla:collage4} provide additional data regarding the percentage of metallicity and age in the stellar profiles of all the Galaxies. These findings further contribute to our understanding of how different DM models influence the stellar properties of galaxies.

\section{ Discussion and conclusions}\label{conclu}
The comparison between the stellar core-halo structures (not the stellar core itself, but the observed inner core and outer halo structure in Figures \ref{figure2} and \ref{CFAMCMC}, first detected in \cite{Pozo:2020}) of $\psi$DM simulations with the CDM predictions reveals intriguing distinctions. In the case of $\psi$DM, the core-halo structure appears to naturally stem from the also observable DM core-halo structure (see the first panel of the top row of figure \ref{all}), whereas for CDM, this structure is not anticipated. Moreover, as seen in figure \ref{all}, it is clear that in DM the N-body simulations exhibit a cuspy DM profile, whereas only $\psi$DM presents solitonic cores. Any core formation in CDM and “WDM” seems to be driven by baryonic physics rather than being directly related to the DM component. Therefore, baryonic physics is the most likely explanation for the transformation of the observed DM cusp in “WDM”--CDM into stellar cores. This notion could be further examined by tracking the stars of each galaxy across different epochs and observing their initial positions to scrutinize in greater detail whether the stars appearing in the outer extended halos were born within their halo or in the outer filamentary structures. This observation, combined with other disparities between CDM predictions and observations, has spurred the investigation of alternative DM candidates such as “WDM” and $\psi$DM. Moreover, the fact that the stellar profiles in each DM scenario follow  the predicted behaviors, with a cuspy profile for CDM and a flat or prominent core structure for $\psi$DM, supports the notion that stars could trace DM. Figure \ref{all} vividly illustrates significant differences in the density profiles of  DM, stars, and gas in each DM scenario. The resulting stellar profile varies depending on the DM model, indicating a cuspy--core discrepancy and core-halo structure. Finally, this work aims to clarify how the core-halo structure can evolve and how definitive it can be to support any dark matter model. It can be observed that the core-halo structure appears in the $\psi$DM dark matter density profiles but never for CDM--“WDM”. This point alone might not be significant enough if it were not for the fact that such core-halo structures, which we are observing for dark matter, also appear in the $\psi$DM Galaxy 1 stellar profile and probably in Galaxy 3 for “WDM” at z=2.23, as can be seen in Figures \ref{figure2} and \ref{CFA“WDM”223}. However, it never appears for CDM. This strongly supports the assertion that the stellar profiles show discrepancies for the three different dark matter models, as we claim in this work.

This underscores the potential significance of studying the properties of stars to achieve a deeper comprehension of the characteristics of DM. Nonetheless, it is also true that in  CDM and “WDM”  the  stellar and DM profiles both appear to shift from a cuspy profile to a cored one when decreasing the redshift, although they never become as flat and prominent as the $\psi$DM predictions, which exhibit a better consistency with the observations \citep{Sanchez:2023}. This transformation from cuspy to cored profiles for  CDM--“WDM” seems to align with the asserted impact of baryonic feedback in generating cores from originally cuspy halos. While in the case of “WDM”, the inner cores are claimed as well \citep{Maccio:2012,Lovell:2014,Bode:2001,Destri:2013}. Feedback induces bursts of star formation and outflows, modifying the DM distribution and mitigating small-scale issues in CDM \citep{Read:2005, Amorisco:2013, Chan:2015}, potentially resolving the cuspy-core problem. Furthermore, the simulations conducted by \citet{Mocz:2019, Mocz:2020} were highly informative, demonstrating that baryonic feedback may not possess sufficient strength to convert cusps into cores for z>6, but it does open up possibilities for lower redshifts. This coincides with our findings where the cusps observed for CDM--“WDM” at z$\simeq$5.56 seem to be alleviated around z$\simeq$2.23.

Certainly, the possible detection of the stellar core-halo structure in “WDM” simulations, despite the absence of a soliton, is a topic of discussion. It seems that the smoothness of the cutoff in the initial dark matter power spectrum, which is likely more pronounced for  “WDM”, may play a role in the formation of the core-halo structure in the stellar profiles. This suggests that the soliton-interference pattern change of regime itself may not be the sole factor determining the presence of the core-halo structure.  Nevertheless, it is extremely important to note that this conclusion is a product of our poor understanding of why the stellar core-halo structure appears for “WDM” (at z = 2.23, Figure \ref{CFA“WDM”223} Galaxy 3) as we thought that it could only appear as a consequence of the expected profile observed in all ideal $\psi$DM dark matter models, featuring an inner flat core fitted with a soliton and an outer characteristic $r^{-3}$ asymptotic profile. Even more analysis is thus required for this topic, and we conclude that it has to be a coincidence due to the absence of such structure for “WDM” dark matter density profiles (see figure \ref{all}).

It is worth noting that the underlying cause of the cutoff differs for “WDM” and $\psi$DM;  “WDM” exhibits a thermal free-streaming cutoff, while $\psi$DM features a quantum mechanical cutoff. However, this distinction does not seem to significantly impact larger scales where the cutoff length scale is imprinted on the profiles. The simulation data presented by \cite{Mocz:2020} demonstrates that only the $\psi$DM core-halo stellar profiles are reliable, due to the presence of this structure in both gas and DM, which is only noticeable in $\psi$DM simulations.

On the other hand, the discrepancies found in the stellar morphology, age, and metallicity suggest the notion of a less concentrated and noncuspy inner halo as a prerequisite for the observed core-halo structure in local dwarfs. This idea gains further support when we observe the higher concentration of stars in CDM--“WDM” models compared to $\psi$DM, which exhibit a lower proportion of stars in their halo centers, indicating that the stellar profiles are more extended. Furthermore, the spatial distribution of metallicity and age indicates that the core-halo structure emerges in galaxies where the greater concentration of stars is not extremely close to the center, a condition that appears only feasible in a filamentary environment, which is challenging to replicate for CDM. Finally, the solitonic stellar imprint is substantiated by the observed asymmetry in the core of G1 in the $\psi$DM scenario, as a consequence of the previously analyzed solitonic random walk \citep{Li:2021, Chiang:2021}, where stars are thought to sense the movement of the soliton and respond accordingly, resulting in the observed asymmetry, in stark contrast to CDM--“WDM” outcomes. This further strengthens the idea of distinct stellar behaviors for each DM model. Quantifying this effect with the deepest JWST images of different dwarfs at various redshifts could yield valuable insights. Additionally, a more detailed analysis of this soliton jumping phenomenon could be carried out with data from lower redshifts, an area we plan to explore in future studies by tracking the motion of stars around the soliton across different redshifts.

The light boson used here, 2.5 $\times 10^{-22}$eV, is consistent with the canonical value based the internal properties of the classical dwarf spheroidal galaxies \citep {Schive:2014, Chen:2017, Pozo:2023}, which has been claimed to be significantly lighter than some independent constraints using the Ly-forest of $>2\times10^{-21}$eV \citep{Irsic:20172}, but it should be noted that the forest estimates are not direct and rely on analogy with “WDM” simulations as they have similar power spectrum suppressions at high frequency (from the de Broglie scale and the free streaming scale, respectively); however, the $\psi$DM interference introduces additional power on small scales that is pervasive, including the low-density regions probed by the Ly-forest, where a wide range of the de Broglie scale is seen in the simulations extending to low momentum in the voids (see Figure S2 of \cite{Schive:2014}). Furthermore, the claimed discrepancy with $\psi$DM is below lines widths of $\simeq 20km/s$ and only at z$>$5, in the steeply declining Jeans turnover where the gas temperature is degenerates with the DM power spectrum, and where spatial temperature variations from AGN and cluster galaxy gas heating sources \citep{Doughty:2023} adds to the forest variance, with widespread metal injection and wide forest gaps  \citep{Yongda:2022}, thus raising the observed power spectrum; however,  all this is too uncertain to confidently model in forest simulations. Hence, reliance on Ly-forest regarding the boson mass of Wave DM is far less direct than our lensing-based estimate \citep{Amruth:2023,Broadhurst:2024}. However, insisting on  smaller scale power can be simply achieved with an additional heavier boson species \citep{Rogers:2021}, as in the generic string axiverse \citep{Arvanitaki:2010}. An additional heaver boson is also motivated by the compact scale of the new UFG class of dwarf galaxy where a boson mass for which a boson mass of $\simeq 2.3\times10^{-21}$eV is indicated dynamically \citep{Pozo:2023}, forming a minority of the DM, most of which is comprised of the canonical $\simeq 10^{-22}$eV boson species. Galaxy formation in this two-boson model has been shown to be viable, capable of forming UFG dwarfs \citep{Luu:2024}.

A lensing based estimate of $>4.4\times10^{-21}$eV is claimed by \citet{Powell:2023}, who argue that their reconstructed source is fully resolved with VLBI, so that sufficiently light $\psi$DM would cause disruption. The breadth of their predicted band of $\psi$DM critical curves is wide, about 25\% of the Einstein radius, whereas a band narrower by a factor of 3 is predicted by Eq. 3, and by comparable lens simulations of $\psi$DM \citep{Chan:2020,Amruth:2023}, which would leave the source unscathed, and hence insensitive to the lighter boson mass we estimate here. 

A tight constraint of $m_b$>3$\times10^{-19}$eV has been claimed dynamically by \cite{Dalal:2022} using two dwarf galaxies Segue I and II. This claim rests mainly on Segue II; the significance of their boson mass estimate relies mainly on the viability of  Segue II, because its velocity dispersion is too small to be detected, unlike the 5.4 Km/s dispersion established for Segue I. This reliance on Segue II is evident in Figure 4 of \cite{Dalal:2022}. However, this dwarf Segue II is widely recognized as an extreme tidal remnant in the careful work of Kirby et al. 2013, and follow-up work where its heavy progenitor origin is reinforced by star orbits\citep{Simon:2019}. Kirby \citep{Kirby:2013} point out that Segue II is an extreme outlier in terms of its high metallicity and low (undetected) velocity dispersion, which they conclude implies a much more massive origin similar to the regular dSph galaxy, and also conclude that Segue II may be the barest remnant of a tidally stripped Ursa Minor-sized galaxy. If this is the case, it is the best example of an ultra-faint dwarf galaxy that became  ultra-faint through tidal stripping. Thus, it is widely believed that Segue II is not representative of the class of low-mass galaxies that \cite{Dalal:2022} would like to use to constrain DM. On the other hand,  Segue I appears to be a bona fide representative of the ultra-faint and dark matter-dominated dwarf with a well-detected velocity dispersion.  A Jeans analysis of this dwarf finds a boson mass of $ \simeq 3 \times 10^{-21}$eV \citep{Pozo:2023}. This is an order of magnitude difference, and  the UFD class as a whole indeed fits well with this higher boson mass. A case for two boson species has been made to encompass both UFD and dSPh classes by some of us \citep{Pozo:2023}. The argument of \cite{Dalal:2022} rests on simulations that seem to underpredict the observed velocity dispersion of 3.5 km/s, so that it is not clear if Segue II alone challenges $\psi$DM. Instead, independent Jeans analysis shows that it is consistent in terms of boson mass with the other UFD galaxies with $\simeq 3 \times 10^{-21}$eV \citep{Pozo:2023} and distinct from the $10^{-22}$eV dominant light boson indicated by the dark matter dominated dSph class \citep{Pozo:2023}. Two such light boson species are arguably well motivated by the axiverse scenario where a discrete mass spectrum of light bosons is generic to string theory \cite{Arvanitaki:2010}, and is the subject of recent multi-boson species simulations \cite{Luu:2024,Chan:2020}.

We conclude that CDM is under relative tension with stellar core-halo observations \citep{Pozo:2020,Pozo:2023}, as it is never detected for any simulated CDM halo. More analysis in different simulations needs to be done in the future to truly support the idea that CDM does not produce these stellar core-halo structures in galaxies. Moreover, there are several compelling reasons to favor $\psi$DM over “WDM”--CDM. First, as highlighted previously, $\psi$DM seems to align better with observations in terms of the presence and structure of core-halo profiles in galaxy halos, as opposed to CDM. Additionally, $\psi$DM predicts a granularity in the lensed halos \cite{2023NatAs...7..736A} that has not been observed in “WDM” simulations. Furthermore, the lack of detection of “WDM” particles in laboratory searches is a substantial drawback for this model. In contrast, $\psi$DM posits light bosons that would necessitate a much larger laboratory scale of 1 kpc for detection, making it challenging to rule out this possibility based solely on laboratory experiments. Overall, the combination of better agreement with observations and the lack of definitive evidence against it makes $\psi$DM a promising candidate for dark matter, and a more viable  alternative to “WDM”--CDM.

\begin{acknowledgements} 
We warmly acknowledge Douglas Finkbeiner for very fruitful conversations. A.P. is grateful for the continued support of the DIPC graduate student program as well as the Center for Astrophysics | Harvard \& Smithsonian for their warm hospitality at the beginning of this project. R.E. acknowledges the support from grant numbers 21-atp21-0077, NSF AST-1816420, and HST-GO-16173.001-A as well as the Institute for Theory and Computation at the Center for Astrophysics. We are grateful to the supercomputer facility at Harvard University where most of the simulation work was done. G.S. is grateful to the IAS at HKUST for their generous support. This work has been supported by the Spanish project PID2020-114035GB-100  (MINECO/AEI/FEDER, UE).  
P.M. acknowledges this work was in part performed under the auspices of the U.S. Department of Energy by Lawrence Livermore National Laboratory under contract DE-AC52-07NA27344, Lawrence
Livermore National Security, LLC.

\end{acknowledgements} 

\section*{Data availability}

The data underlying this article will be shared on reasonable
request to the corresponding author.

\bibliographystyle{aa} 
\bibliography{apssamp}

\newpage
\begin{appendix}

\onecolumn

\section{Stellar Morphology at $z=2.23$}
Here we present the equivalent plots for data at $z=5.56$ from the main text, for $z=2.23$. This supplementary analysis further enriches our understanding of the core-halo structure in the system under investigation. It is valuable to explore how these parameters interact and correlate at different redshifts, providing a more comprehensive view of the underlying dynamics. Figures \ref{starsmorpho4}, \ref{age2} and \ref{starsmorpho6}.

\begin{figure}[H]
	\centering
	\includegraphics[width=0.9\textwidth,height=20cm]{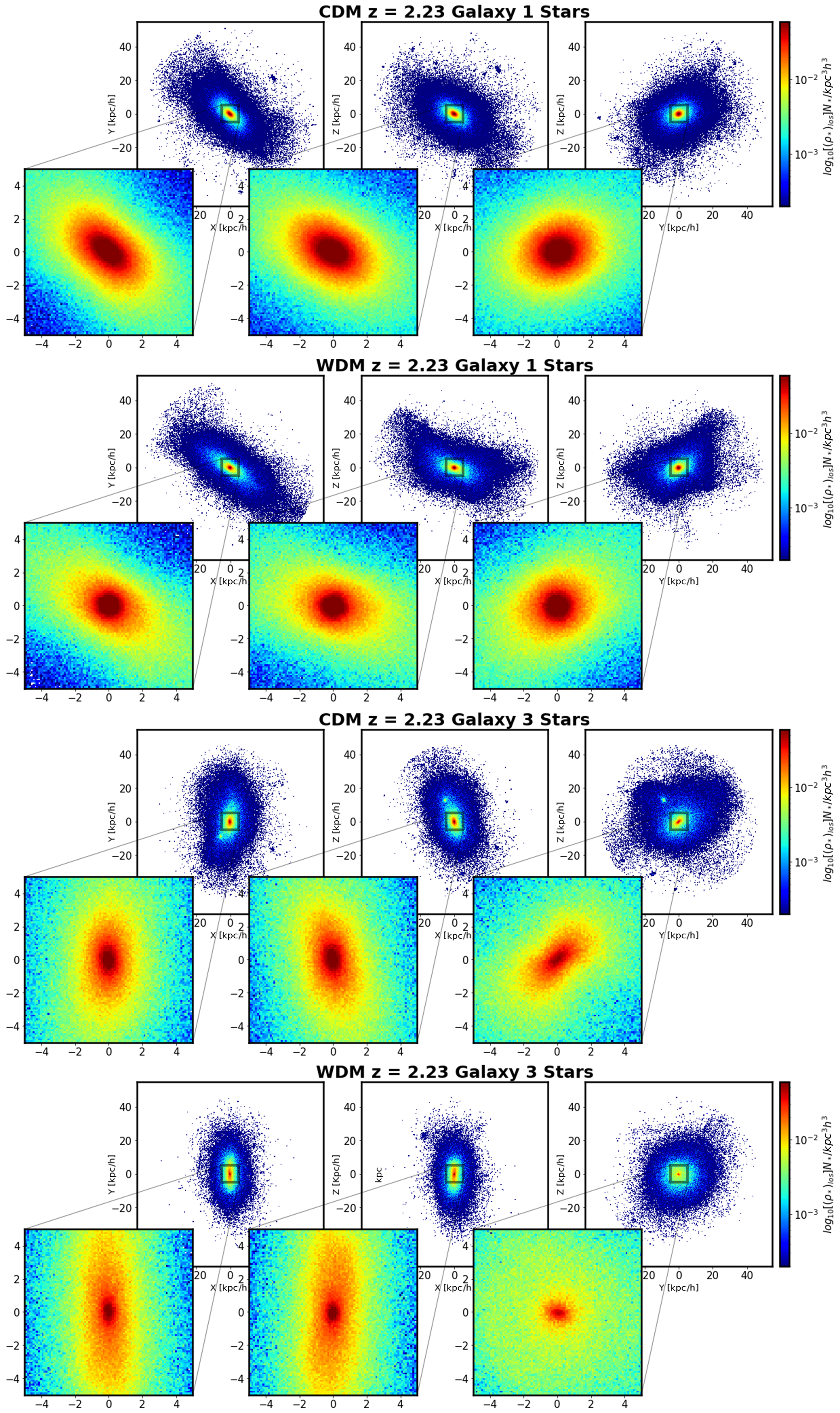}
    
	\caption{ Logarithm of the projected number density of stellar particles in each simulated galaxy. The row headers indicate the halo number, the DM model, and the redshift. Additionally, we  zoom in on the central regions of the galaxies for a closer view. The data is represented in comoving units.}\label{starsmorpho4}
\end{figure}

\begin{figure}[H]
	\centering
	\includegraphics[width=0.9\textwidth,height=20cm]{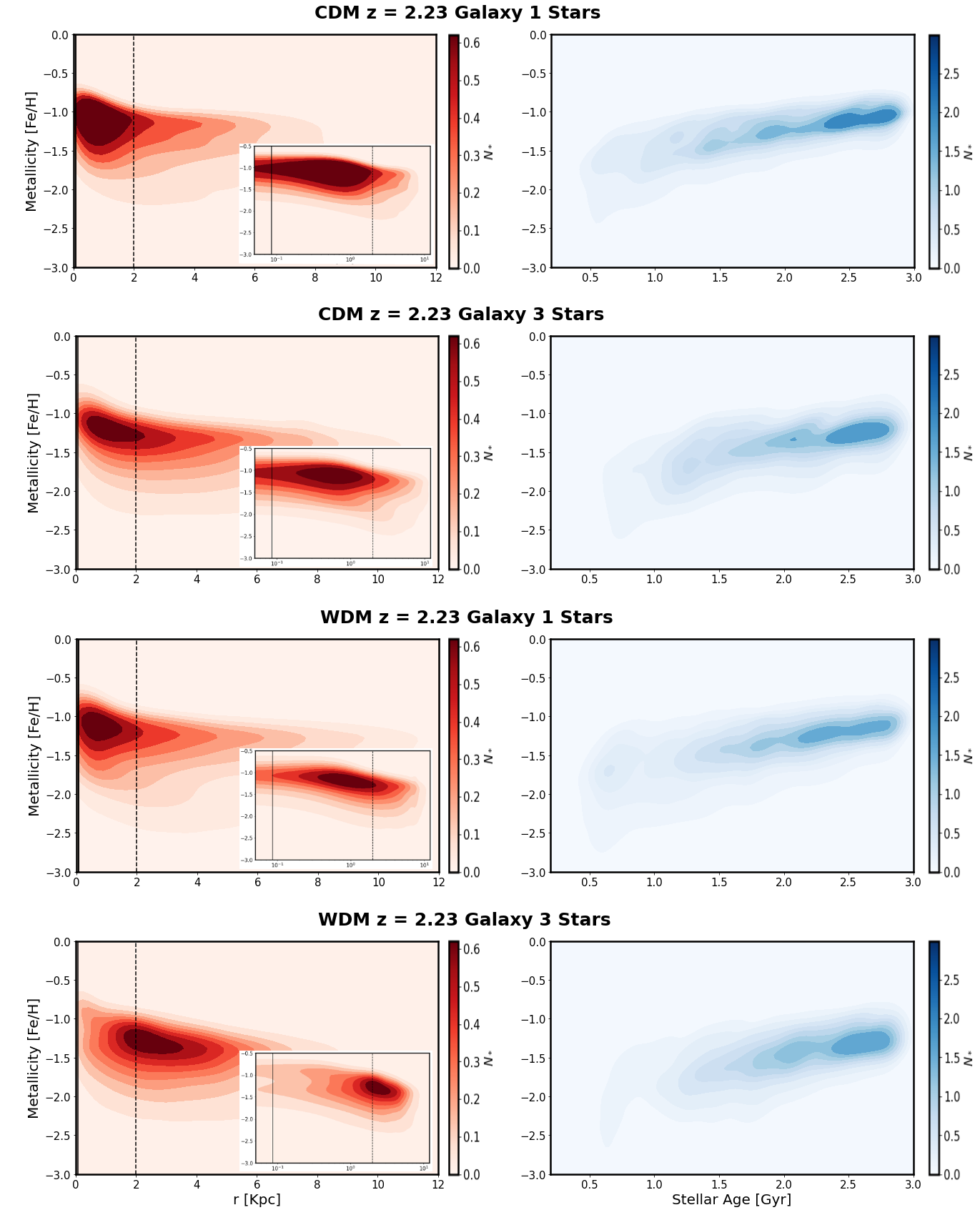}
	\caption{Left panel: 2D correlation of the stellar distance vs. stellar metallicity. It is evident that metal-rich stars tend to be located closer to the galactic center. However, there is a notable offset for “WDM”$_{G3}$, suggesting a unique distribution pattern for this specific galaxy. The small subpanels provide a logarithmic scale representation for enhanced visualization. The vertical black line signifies the comoving limit, while the dashed line represents the ideal size of a core-halo structure.  Right panel: 2D correlation of stellar age vs. metallicity. The negative Spearman correlation coefficient indicates that older stars are more likely to be metal-poor, while younger stars tend to be more metal-rich. This relationship holds across all DM models. }\label{age2}
\end{figure}

\begin{figure}[H]
	\centering
	\includegraphics[width=0.9\textwidth,height=20cm]{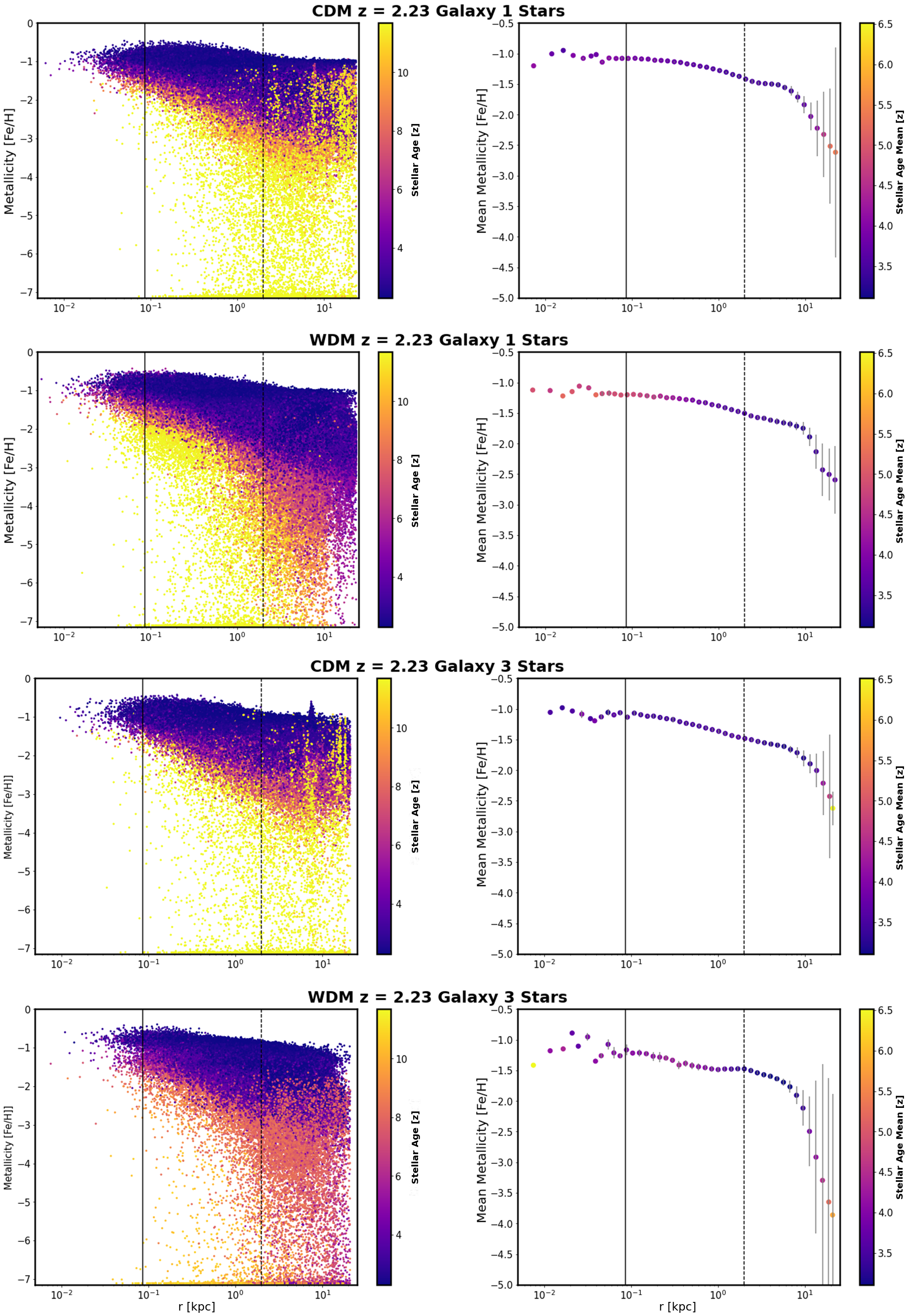}
	\caption{Left panel: 2D correlation of  stellar distance vs. stellar metallicity. It is evident that metal-rich stars are predominantly located closer to the galactic center. This trend holds across all DM models, with a slight offset observed for $\psi$DM. The data points are color-coded based on their stellar age, providing additional context.  Right panel: 2D correlation of stellar spatial distribution vs. mean metallicity. The negative Spearman correlation coefficient indicates that older stars tend to be more metal-poor, while younger stars are more likely to be metal-rich. The data points are color-coded based on their mean stellar age, offering further insights. The vertical black line indicates the comoving limit, while the dashed line represents the ideal size of a core-halo structure. }\label{starsmorpho6}
\end{figure}

\newpage

\section{Tables with metallicity, age and spatial distribution of stars}
Here we present a summary in two tables ( Tables \ref{tabla:collage3} and \ref{tabla:collage4}) of the metallicities, ages, and spatial distribution of the stars in the three simulated halos. These tables are interesting for quantitatively observing the distribution of stars in terms of age, metallicity, and spatial location within both the cores and halos of the simulated Galaxies across the three dark matter models.

\begin{table*}[htbp]
\caption{Summary of the metallicity of the total number of stars in the first 2 kpc of every halo.}
	\centering
\begin{tabular}{|c|c|c|c|c|c|}
\hline
 Halos& $N_{*2{\rm kpc}}$ & (-2<[Fe/H]<0)  &(-4<[Fe/H]<-2) &(-6<[Fe/H]<-4)&(-8<[Fe/H]<-6) \\
& &$\%$ &$\%$&$\%$&$\%$ \\
\hline
$\psi$DM$_{G1}$ &12340 & 93.12 & 4.91 & 0.71 & 1.26 \\
\hline
$\psi$DM$_{G2}$ &1950 &80.62&14.82&1.44&3.12  \\
\hline
$\psi$DM$_{G3}$  &1550 &77.22& 16.91 & 1.68 & 4.19\\
\hline
CDM$_{G1}$  & 21206& 95.62 & 3.61 & 0.36 & 0.41  \\
\hline
CDM$_{G2}$ & 7572&88.55&9.17&1.12& 1.16\\
\hline
CDM$_{G3}$  &3417 &83.79&12.38&1.82&2.01  \\
\hline
“WDM”$_{G1}$ &28328 &89.48 & 8.43 & 0.76 & 1.33 \\
\hline
“WDM”$_{G2}$  &3793 &79.25&14.79&2.14& 3.82 \\
\hline
“WDM”$_{G3}$  &5122 & 78.21 & 16.34 & 1.44 & 4.01 \\
\hline
\hline
\hline
\end{tabular}
\tablefoot{Column 1: Simulated Dwarf individual name, Column 2: Number of stars inside the first 2kpc  $N_{*2{\rm kpc}}$,  Column 3: Percentage of stars in column 2 with a [Fe/H] between 0 and -2, Column 4: Percentage of stars in column 2 with a [Fe/H] between -2 and -4, Column 5: Percentage of stars in column 2 with a [Fe/H] between -4 and -6, Column 6: Percentage of stars in column 2 with a [Fe/H] between -6 and -8. Data for halos at z=5.56. }
\label{tabla:collage3}
\end{table*}

\begin{table*}[htbp]
\caption{Summary of the age and spatial distribution of the total number of stars in the first 2 kpc of every halo.}
\centering
\begin{tabular}{|c|c|c|c|c|c|c|c|c|}
\hline
Halos& $N_{*2{\rm kpc}}$ & $N_{*2{\rm kpc}}/N_{*{\rm Total}} $& $N_{*0.5{\rm kpc}}/N_{*2{\rm kpc}} $& $N_{*0.5{\rm kpc}}/N_{*{\rm Total}}$ & $(0<z<3)$   & $(3<z<5)$ & $(5<z<8)$ & $(8<z<12)$ \\
& &$\%$&$\%$&$\%$&$\%$ &$\%$&$\%$&$\%$ \\
\hline
$\psi$DM$_{G1}$ &12340 & 11.57 &8.02&0.92&48.72 & 37.41 & 10.46 & 3.41 \\
\hline
$\psi$DM$_{G1}$ &1950 & 5.66 & 2.82& 0.16& 72.10  & 16.41 & 7.23& 4.26 \\
\hline
$\psi$DM$_{G1}$  &1550 &5.10&2.89&0.15& 64.45 & 27.36 & 8.19 & 0\\
\hline
CDM$_{G1}$  & 21206&13.21&5.58&0.74& 48.60 & 37.27 & 7.75 & 6.38  \\
\hline
CDM$_{G2}$  & 7572&24.57&8.24&2.02& 29.32 & 30.78 & 14.05 & 25.85\\
\hline
CDM$_{G3}$ &3417 &7.99&10.87&0.87& 35.91 & 24.00  & 9.51 &  30.58 \\
\hline
“WDM”$_{G1}$  &28328 &24.56&13.75&3.37& 38.04  & 36.55 & 12.05 & 13.36 \\
\hline
“WDM”$_{G2}$   &3793 &10.18&6.14&0.63& 63.88 & 20.43 & 6.52 & 9.17 \\
\hline
“WDM”$_{G3}$ &5122 &16.08&9.31&1.50& 49.57  & 38.54 & 11.21 & 0.68 \\
\hline
\hline
\hline
\end{tabular}
\tablefoot{Column 1: Simulated Dwarf individual name, Column 2: Number of stars inside the first 2kpc  $N_{*2{\rm kpc}}$,  Column 3: Percentage of stars in column 2 from the total amount of stars, Column 4: Percentage of stars in column 2 located inside the first 0.5kpc, Column 5: Percentage of stars inside the first 0.5kpc from the total number of stars, Column 6: Percentage of stars with an age redshift between 0 and 3, Column 7: Percentage of stars with an age redshift between 3 and 5, Column 8: Percentage of stars with an age redshift between 5 and 8, Column 6: Percentage of stars with an age redshift between 8 and 12. Data for halos at z=5.56. }
\label{tabla:collage4}
\end{table*}

\newpage
\section{Corner Plots from MCMC fit}
Here we present the resulting corner plot of the performed MCMC fit corresponding to Figure \ref{CFAMCMC}. Figure \ref{cornerCFA}.

\begin{figure*}[htbp]
	\centering
	\includegraphics[width=1\textwidth,height=11cm]{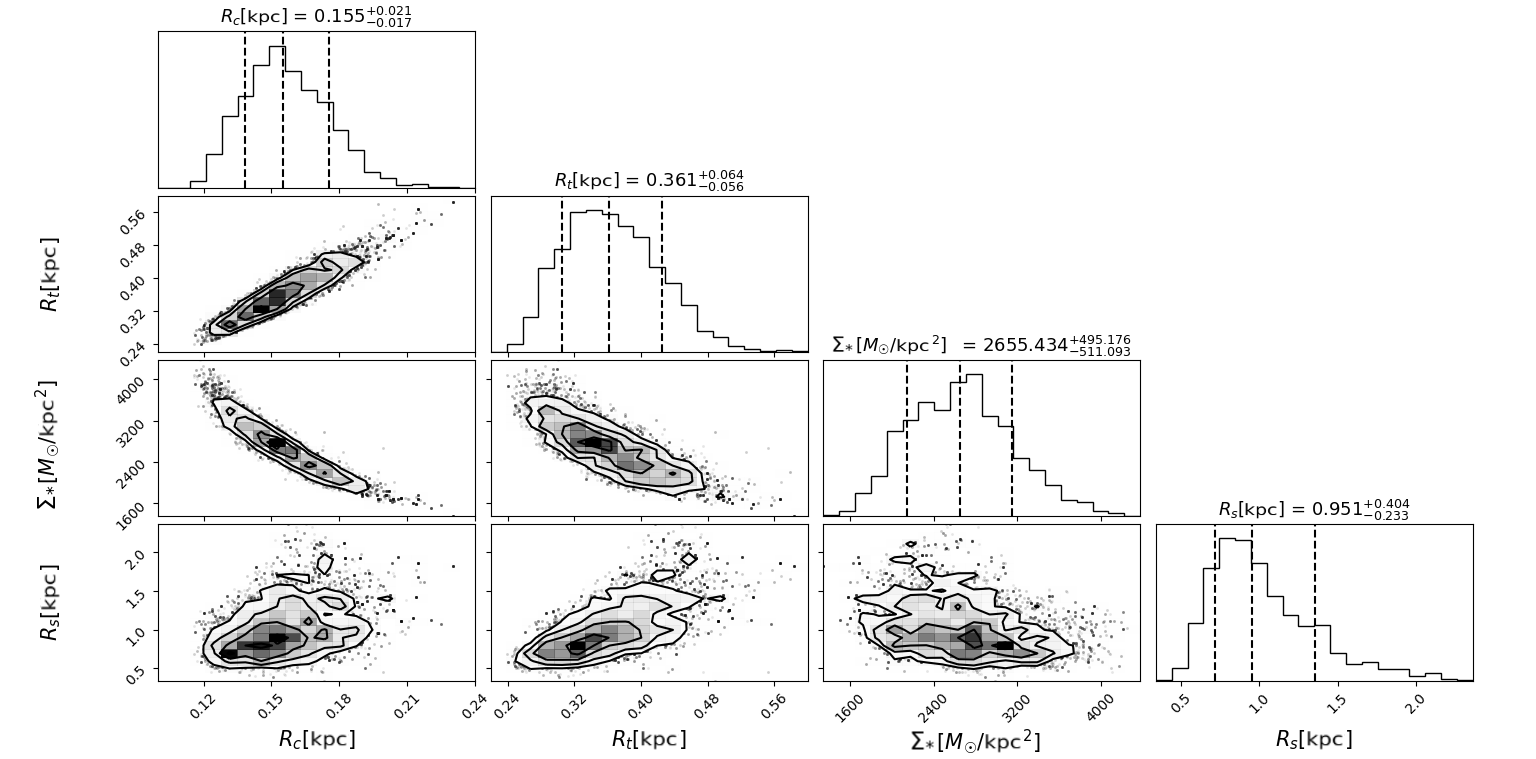}
	\caption{Respective corner plot of  figure \ref{CFAMCMC}. This figure displays the correlated distributions of the free parameters. The core radius and transition radius are particularly well defined, even with the Gaussian input priors, indicating a reliable result. The contours in the plot represent the 68\%, 95\%, and 99\% confidence levels, providing a visual representation of the parameter uncertainties. The best-fit parameter values, along with their respective errors, are indicated by the dashed black lines. These values are also listed in Table \ref{tabla:collage} for easy reference.}\label{cornerCFA}
\end{figure*}

\label{lastpage}
\end{appendix}
\end{document}